\documentclass[final,authoryear,3p,times]{elsarticle}

\usepackage{booktabs}
\usepackage{amssymb}
\usepackage{amsmath}
\usepackage{cases}
\usepackage{mathptmx}
\usepackage[nonumberlist, toc, acronym]{glossaries}
\usepackage{acronym} 
\usepackage{array}

\usepackage[tight,hang]{subfigure}
\usepackage{nomencl}
\usepackage{ifthen}

\renewcommand{\nomgroup}[1]{%
\ifthenelse{\equal{#1}{A}}{}{%
\ifthenelse{\equal{#1}{G}}{\vspace{5mm} \item[\textit{\large{Greek Letters}}]}{%
\ifthenelse{\equal{#1}{Z}}{\vspace{5mm} \item[\textit{\large{Abbreviations}}]}{%
\ifthenelse{\equal{#1}{S}}{\vspace{5mm} \item[\textit{\large{Subscripts}}]}{%
\ifthenelse{\equal{#1}{M}}{\item[\textit{\large{Mathematical Symbols}}]}
{}
}
}
}
}
}

\makenomenclature

\begin{document}

\begin{frontmatter}


\title{A stability condition for turbulence model: From EMMS model to EMMS-based turbulence model}


\author[label1]{Lin Zhang}
\author[label1,label2]{Xiaoping Qiu}
\author[label1]{Limin Wang \corref{cor1}}
\ead{lmwang@home.ipe.ac.cn} \cortext[cor1]{Corresponding author.
Tel.: +86 10 8254 4942; fax: +86 10 6255 8065.}
\author[label1]{Jinghai Li}

\address[label1]{The EMMS Group, State Key Laboratory of Multiphase Complex Systems, Institute of Process Engineering, Chinese Academy of Sciences, Beijing 100190, China}
\address[label2]{University of Chinese Academy of Sciences, Beijing 100049, China}

\begin{abstract}

The closure problem of turbulence is still a challenging issue in turbulence modeling. In this work, a stability condition is used to close turbulence. Specifically, we regard single-phase flow as a mixture of turbulent and non-turbulent fluids, separating the structure of turbulence. Subsequently, according to the picture of the turbulent eddy cascade, the energy contained in turbulent flow is decomposed into different parts and then quantified. A turbulence stability condition, similar to the principle of the energy-minimization multi-scale (EMMS) model for gas-solid systems, is formulated to close the dynamic constraint equations of turbulence, allowing the inhomogeneous structural parameters of turbulence to be optimized. We call this model the `EMMS-based turbulence model', and use it to construct the corresponding turbulent viscosity coefficient. To validate the EMMS-based turbulence model, it is used to simulate two classical benchmark problems, lid-driven cavity flow and turbulent flow with forced convection in an empty room. The numerical results show that the EMMS-based turbulence model improves the accuracy of turbulence modeling due to it considers the principle of compromise in competition between viscosity and inertia.

\end{abstract}


\end{frontmatter}


\section{Introduction}
\label{}
Turbulence is one of the most important unresolved problems in classical physics (Feynman et al., 1963). With the rapid increase of computer capability, numerical simulation of turbulence has attracted considerable attention from researchers and engineers. Direct numerical simulation (DNS) (Moin and Mahesh, 1998) is the most fundamental and accurate numerical approach to simulate turbulence. However, the spatial and temporal scales of DNS are limited by both the Kolmogorov and turbulence time scales, which make it computationally expensive. Compared with DNS, large eddy simulation (LES) (Meneveau and Katz, 2000) shows some promise, although it is also computationally expensive. For example, theoretically DNS will be able to simulate a whole generic aircraft configuration in 2080, while LES will be able to with 90\% of the scales resolved in 2045 (Spalart, 2000). This is because the computational time of DNS and LES increases rapidly as the Reynolds number increases. Therefore, at present there is only one way to solve engineering turbulence problems; namely, through solution of the Reynolds-averaged Navier-Stokes (RANS) equations with the aid of turbulence models (Wilcox, 1998), which is known as the turbulence model theory. Although the computational cost of this method is much lower than those of DNS and LES, turbulence models derived from empirical relationships and experimental data simulate turbulent flows with lower accuracy. Thus, to simulate turbulence better, especially related to practical engineering problems, it is important to improve the current turbulence model.

In gas-solid riser flows, meso-scale structures existing in the form of particle clusters or aggregates have considerable effects on flow behavior. To reasonably describe these meso-scale structures and accurately model global behaviors, eight structural parameters are introduced and then interrelated by six equations for conservation of mass and momentum. However, this does not allow closure, so a stability criterion is proposed to define the steady state using a compromise between different dominant mechanisms; that is, the energy-minimization multi-scale (EMMS) model (Li, 1987; Li and Kwauk, 1994; Li et al., 1988). This model, which is equivalent to the turbulence model for single-phase turbulent flow, has successfully resolved the meso-scale structures of gas-solid systems (Ge and Li, 2002; Li et al., 1999a; Li et al., 1990; Liu et al., 2011; Lu et al., 2009; Naren et al., 2007; Nikolopoulos et al., 2010a; Nikolopoulos et al., 2010b; Qi et al., 2007; Wang et al., 2008; Wang and Li, 2007; Yang et al., 2003; Yang et al., 2004). In addition, the principles of the EMMS model have also been tested for both gas-liquid systems (Chen et al., 2009a, b; Ge et al., 2007; Yang et al., 2010; Yang et al., 2007; Yang et al., 2011; Zhao, 2006) and single-phase turbulent flow (Li et al.,1999b).

In single-phase turbulent flow, Li et al. (1999b) identified a compromise mechanism between inertial and viscosity effects, and analysis of single-phase turbulent flow in a pipe allowed them to propose a stability condition for turbulence. Wang et al. (2007; 2008) subsequently analyzed the multi-scale structure and energy dissipation process of turbulence to improve this stability condition. At the same time, they also verified the rationality of the improved stability condition by simulating the flow around a cylinder using a particle-based DNS method. The key to establishing a rational turbulence model lies in understanding the structure of turbulence in computational grids, which is strongly related to a stability condition. In other words, it is impossible to develop a better model without including a stability condition. However, none of the current turbulence models consider this point or include a stability condition. For example, the zero-equation model (Pope, 2000) is the simplest turbulence model, which obeys the Boussinesq assumption and uses an algebraic equation (i.e., an empirical formula) for mean velocity and geometric length scale to compute the turbulent viscosity coefficient. Another example is the one-equation model separately proposed by Kolmogorov (1942) and Prandtl (1945); its core idea is to use root mean square (RMS) turbulent kinetic energy as the velocity scale of turbulent fluctuation and then to construct, model and solve the differential equation of turbulent kinetic energy, where length scale is specified algebraically based on the mean flow. The two-equation model (Pope, 2000) uses a supplementary equation, namely the turbulent energy dissipation rate equation successively developed by Chou (1945), Davidov (1961), Harlow \& Nakayama (1968), and Jones \& Launder (1972), to determine the length scale rather than specifying it empirically, as well as using turbulent kinetic energy to determine the velocity scale.

To improve the current turbulence models, a turbulence stability condition should be introduced. As soon as turbulence sets in, multiple mechanisms lead to its typical nonlinear, non-equilibrium dissipative characteristics (Prigogine 1967). Because of the coexistence of two different mechanisms (viscosity and inertia) in a single-phase flow system, the variational criterion of single-phase flow cannot be represented only by the extremum tendency for the viscosity effect or by that for the inertia effect; instead, compromise between these two competing mechanisms plays an important role in system stability. Although considerable inertial dissipation exists in turbulent flow, viscous dissipation $W_{\rm{\nu}}$ maintains the same inherent tendency as in laminar flow; that is, it is minimized (i.e., ${W_{\rm{\nu}}} \to \min$) even though its minimum is subject to inertia. During the study of gas-solid two-phase flow (Li and Kwauk 1994; Li et al. 1996; Li et al. 1998), the appearance of a dissipative flow structure was shown to maximize total dissipation. If single-phase turbulent flow is considered to consist of many turbulent eddies and the interaction of an eddy with its surroundings is similar to that between gas and particles, the dissipative behavior of these two systems is similar. Therefore, a tendency to maximum total dissipation $W_{\rm{T}}$ in turbulent flow could be assumed (i.e., $W_{\rm{T}} \to \max$). This tendency is, however, subject to the simultaneous presence of viscosity, and is not realized exclusively. Based on the above consideration, in real turbulent flow, neither viscous nor inertial effects can dominate a system, so they fail to realize their respective tendencies exclusively; instead, they play a joint role to achieve system stability by compromising with each other (i.e.,  ${W_{\rm{\nu}} } \to \min \left| {_{{W_{\rm{T}}} \to \max }} \right.$). In physics, fluid inertial force is the force exerted on a fluid by temporal fluctuation. Therefore, maximization of turbulent (fluctuating) energy dissipation $W_{\rm{te}}$ (i.e. $W_{\rm{te}} \to \max$) is a better qualitative and quantitative indicator for the effect of inertia than $W_{\rm{T}}$ (Wang et al., 2007). As a result, the stability condition for the evolution of turbulent flow can be characterized by  ${W_{\rm{\nu}} } \to \min \left| {_{{W_{\rm{te}}} \to \max }} \right.$.

The turbulence stability condition is a more physically reliable and accurate alternative to improve current turbulence models because it is a control mechanism of objective existence rather than being derived from empirical relationships and experimental data. In this paper, based on the two-phase concept of turbulence and estimation of the difference in density between turbulent and non-turbulent eddies (i.e., the laminar fluid), the structure of turbulence is separated and some constraint equations are constructed with basic principles such as mass conservation and force balance. Subsequently, the energy contained in turbulence is separated into different parts according to the homogeneous isotropic turbulence theory (Batchelor, 1982; Pope, 2000), and each energy component is also quantified. The turbulence stability condition is then introduced to close the constructed constraint equations, allowing the inhomogeneous structural parameters of turbulence to be optimized. Finally, by introducing these optimized parameters into the current turbulence model, the `EMMS-based turbulence model', which considers both flow structure and the turbulence stability condition, is proposed. This new turbulence model can effectively represent the coexistence of laminar and turbulent regions in turbulence on each mesh (grid), rather than the a priori assumption that all complex flows are fully turbulent flows as in traditional turbulence models.

Here, we develop a new turbulence model that considers both flow structure and the turbulence stability condition; i.e., the EMMS principle of turbulence. The rest of this paper is organized as follows: Section 2 presents the EMMS-based turbulence model including some preliminary knowledge, constraint equations of turbulent eddies, energy decomposition and its quantification in turbulence, the turbulence stability condition, the physical basis and a summary of the EMMS-based turbulence model. Section 3 discusses the results obtained using the EMMS-based turbulence model, and the accuracy of numerical examples is improved using the EMMS-based turbulence model. Some conclusions are drawn in Section 4.

\section{The EMMS-based turbulence model}
\subsection{Preliminary knowledge}
\label{}
\begin{itemize}
  \item \textit{A physical picture of the turbulent eddy cascade} (Pope, 2000). The turbulent eddy cascade can be described as follows: the external force acting on a fluid sustains the motion of large-scale eddies, which are unstable and break up to produce smaller ones. These smaller turbulent eddies undergo a similar breakage process to produce even smaller turbulent eddies. The energy loss at each hierarchy of turbulent eddies can be divided into two parts (Xu, 1986): one is the energy transferred at each step from the large turbulent eddies to the smaller ones through their breakage process, and the other is the energy dissipated into heat by the molecular viscosity of the fluid, which occurs inside the turbulent eddies as they break up. Large turbulent eddies have a very high Reynolds number, which reflects the small effect of viscosity, so the transferred energy is greater than that dissipated. This situation is reversed for smaller turbulent eddies. Because external energy is only supplied to the large turbulent eddies, an energy cascade is formed (Pope, 2000): the external energy injected into large turbulent eddies is sequentially transferred to increasingly smaller ones until it is dissipated into internal energy. If external energy is supplied continuously to the large turbulent eddies or they have sufficient stored energy, energy balance may be possible; namely, the input energy equals the output energy, so the fluid motion is steady.\\
 	
  \item  \textit{The two-phase concept of single-phase turbulence.} The ``conditional sampling'' technique (Antonia et al., 1975; Shepherd and Moss, 1982) can be used to discriminate turbulent and non-turbulent zones of flow. This allows turbulence to be considered as the combination of the motions of two fluids, defined as the ``turbulent fluid'' and ``non-turbulent fluid'', along with their interaction (see the two-phase concept of turbulence in Fig. 1). These two fluids coexist in one space, either sharing the space or occurring in the space with their own probability. The fluids are treated as two interpenetrating continua (Fan, 1988; Spalding and Malin, 1984), so turbulent fluid can become  non-turbulent fluid by energy dissipation, and non-turbulent fluid can become turbulent through the entrainment of turbulent fluid. At each location, the amount of each phase present is characterized by its volume fraction, and the volume fraction of turbulent fluid is interpreted as the intermittency factor of turbulence in that position (Fan, 1988; Spalding and Malin, 1984). Overall, this concept separates the structure of turbulence, and allows us to consider the structure of turbulence to help improve the current turbulence model.\\

  \item \textit{Estimation of the difference in density between turbulent and non-turbulent eddies.} The main characteristic of turbulent flow is the existence of many turbulent eddies of different size. Trolinger et al. (2002) noted that turbulent eddies have a great influence on the transient density distribution of flow. They estimated that the density of a turbulent eddy region with concentrated vorticity was less than that of the surrounding fluid (i.e., the non-turbulent fluid), so the density of fluid between adjacent turbulent eddies was relatively high.
\end{itemize}

\begin{figure}[htb]
  \centering
  \includegraphics[width=0.8\textwidth]{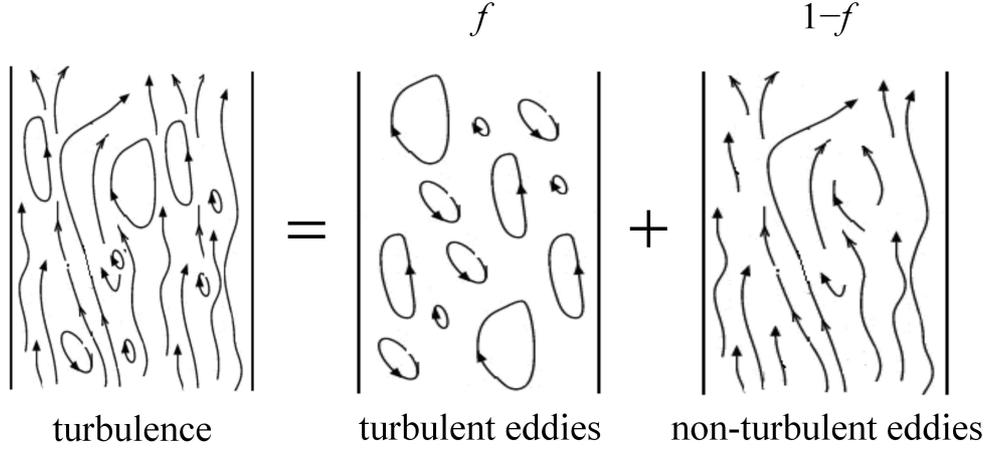}
  \caption{The two-phase concept of turbulence.}\label{fig1}
\end{figure}

\subsection{The constraint equations of turbulent eddies}
In general, the inhomogeneous flow structure of turbulence takes on the existence of turbulent eddies of different size as well as the volume fraction of turbulent eddies in the flow. The size and volume fraction of turbulent eddies are influenced by the operating conditions, which determine the characteristics of the whole fluid flow. Therefore, the equivalent diameter and volume fraction of turbulent eddies can be regarded as the inhomogeneous structural parameters that describe the characteristics of turbulence. Specifically, turbulence as a whole can be characterized by the following inhomogeneous structural parameters: the equivalent diameter of turbulent eddies $d_\mathrm{e}$, the volume fraction of turbulent eddies $f$, the superficial velocity of turbulent eddies $U_\mathrm{e}$ and the superficial velocity of laminar flow $U_\mathrm{l}$.

Based on the estimation of the difference in density between turbulent and non-turbulent eddies, the density of turbulent eddies $\rho\,_\mathrm{e}$ is less than that of the laminar component $\rho\,_\mathrm{l}$. For simplicity, it is assumed that $\rho\,_\mathrm{e}=0.99 \rho\,_\mathrm{l}$. The coefficient will be discussed further later. Moreover, it is also assumed that the input energy is equal to the output energy; namely, the fluid motion is steady. Therefore, for this steady state, the turbulent eddies are balanced by forces. That is, the drag force equals the buoyant force:
\begin{equation}
  \frac{\pi}{6}d_{\mathrm{e}}^{3}\left(\rho\,_\mathrm{l}-\rho\,_\mathrm{e}\right)g=C_\mathrm{D,eddy}\frac{\pi}{4}d_{\mathrm{e}}^{2}\frac{\rho\,_\mathrm{l}}{2}u_\mathrm{s}^2
\end{equation}
where $g$ is gravity acceleration, and $u_\mathrm{s}$  is the slip velocity between turbulent eddies and the laminar component. That is to say,
\begin{equation}
  u_\mathrm{s}=\frac{U_\mathrm{e}}{f}-\frac{U_\mathrm{l}}{1-f}
\end{equation}
Additionally, in Eq. (1), $C_\mathrm{D,eddy}$ is the drag coefficient of turbulent eddies, and the specific expressions can refer to those for a bubble (Lo et al., 2000; Yang et al., 2007); namely,
\begin{equation}
  C_\mathrm{D,eddy}=C_\mathrm{D0,eddy}\left(1-f\right)^{4}
\end{equation}
where
\begin{equation*}
  C_\mathrm{D0,eddy}=\frac{4}{3}\frac{g d_\mathrm{e}}{U_\mathrm{T}^2}\frac{\rho_\mathrm{l}-\rho_\mathrm{e}}{\rho_\mathrm{l}}
\end{equation*}
\begin{equation*}
  U_\mathrm{T}=\frac{\mu _{\,\mathrm{l}}}{\rho _{\,\mathrm{l}} d_\mathrm{e}}M_{\mathrm{o}}^{\,-0.149}\left(J-0.857\right)
\end{equation*}
\begin{equation*}
  M_\mathrm{o}=\frac{g \mu_{\,\mathrm{l}}^{4}\left(\rho_{\,\mathrm{l}}-\rho_{\,\mathrm{e}}\right)}{\rho_{\,\mathrm{l}}^{2}\sigma^3}
\end{equation*}
\begin{equation*}
  J = \left\{ \begin{split}
&0.94{H^{\,0.757}}& 2 < H \le 59.3\\
&3.42{H^{\,0.441}}& H > 59.3
\end{split} \right.
\end{equation*}
\begin{equation*}
  H=\frac{4}{3}E_\mathrm{o} M_\mathrm{o}^{\,-0.149} \left(\frac{\mu_{\,\mathrm{l}}}{0.0009}\right)^{-0.14}
\end{equation*}

\begin{equation*}
  E_\mathrm{o}=\frac{d_{\mathrm{e}}^{2} \left(\rho_{\,\mathrm{l}}-\rho_{\,\mathrm{e}}\right) g}{\sigma}
\end{equation*}

Eq. (1) clearly shows that once the operating conditions (i.e., $U_\mathrm{e}$ and $U_\mathrm{l}$) are given and the equivalent diameter of turbulent eddies $d_\mathrm{e}$ is appointed, the volume fraction of turbulent eddies $f$ will be determined exclusively for a turbulent system. Because the inlet velocity is
\begin{equation}
  U_\mathrm{in}=\left(1-f\right)U_\mathrm{l} + fU_\mathrm{e}
\end{equation}
once the inlet velocity $U_\mathrm{in}$  is given, the parameters $U_\mathrm{e}$, $U_\mathrm{l}$, $d_\mathrm{e}$ and $f$ can be determined uniquely. For simplicity, we assume $U_\mathrm{e}$ and $U_\mathrm{l}$ to be given first and then determine $d_\mathrm{e}$ and $f$.

\subsection{Energy decomposition and its quantification in turbulence}
Similar to gas-solid and gas-fluid systems, turbulence is also a typical complex system, exhibiting the common characteristics of multi-scale structure. The multi-scale structures of the two former systems have already been preliminarily analyzed (Li and Kwauk, 1994; Yang et al., 2007; Zhao, 2006). However, the same work for turbulence has not been reported. In view of this current research situation, we decided to analyze the multi-scale structure of turbulence according to the turbulent eddy cascade. As mentioned above, when the input energy is equal to the output energy, the motion of turbulence is steady. Here, we only analyze the multi-scale structure of turbulence in this flow state. Compared with laminar flow, the main characteristic of turbulence is the existence of turbulent eddies of different size as well as their breakage. Therefore, we can analyze the multi-scale structure of turbulence according to the different size of the turbulent eddies. The mode of energy dissipation varies with the size of a turbulent eddy. The energy contained in turbulence can be classified as follows (a diagram showing energy decomposition in turbulence is shown in Fig. 2):

\begin{figure*}[htb]
  \centering
  \includegraphics[width=0.85\textwidth]{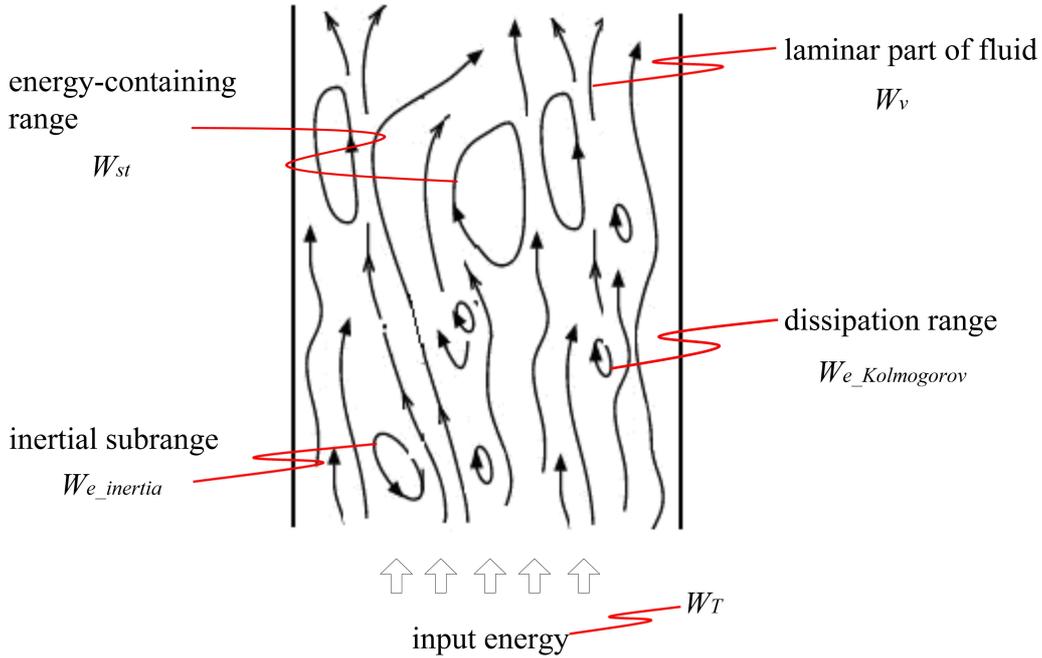}
  \caption{Diagram showing energy decomposition in turbulence.}\label{fig2}
\end{figure*}

\begin{itemize}
\itemsep=1.5ex
  \item {\it Total energy of turbulence} $W_\mathrm{T}$. To maintain the fluid in a turbulent state, external energy needs to be supplied continuously. The energy supplied to generate turbulence in a volume of fluid per unit mass and per unit time is called the total energy of turbulence and labeled as $W_\mathrm{T}$.
  \item {\it Energy storage of energy-containing eddies} $W_{\mathrm{st}}$. Most energy of turbulence is stored in turbulent eddies of large size (although normally not the largest size of turbulent eddies). The energy contained in smaller turbulent eddies is much lower than that in larger ones because of strong viscous dissipation. Therefore, the large turbulent eddies that store most of the energy of turbulence are called energy-containing eddies, and their corresponding size is called the energy-containing range. The energy-containing eddies obtain energy directly from the average motion of fluid or the generating device of turbulence (e.g., an oscillating grid). Without doubt, the motion of these turbulent eddies is directly related to the external conditions, which leads to the generation of non-isotropic turbulent eddies. Moreover, it has been pointed out previously that for large turbulent eddies, there is only transfer energy without dissipated energy (Pope, 2000). Therefore, the transfer energy that is transferred to smaller turbulent eddies is just the energy stored in the energy-containing eddies. The energy-containing eddies are non-isotropic and the violent oscillations of their surfaces can be regarded as a measure of the energy stored in the energy-containing eddies, which is called the energy storage of energy-containing eddies and labeled as $W_{\mathrm{st}}$.
  \item {\it Energy dissipation of the inertial subrange} $W_{\mathrm{e\_\,inertia}}$. Once the energy is stored in the energy-containing eddies, it begins to be transferred to smaller turbulent eddies through breakage of the energy-containing eddies. When a certain hierarchy of turbulent eddies is reached, there is not only transfer energy but also dissipated energy. This is because the Reynolds number decreases as the size of the turbulent eddies decreases, which means that the inertia effect becomes progressively weaker and the viscosity effect becomes stronger. Specifically, the energy loss of turbulent eddies of size equal to or smaller than that of this certain hierarchy of turbulent eddies can be divided into two parts: one is the energy transferred to smaller turbulent eddies through the breakage of the current turbulent eddy, and the other is the energy dissipated into internal energy by molecular viscosity. The latter mainly occurs inside the current turbulent eddy as it breaks up. Furthermore, we think that the inertial subrange starts from this hierarchy, and holds until another certain hierarchy of turbulent eddies is reached, from which there is only the energy dissipated by molecular viscosity but no energy transferred by inertia effect. In the inertial subrange, the first part of energy loss (i.e., the transfer energy) for each hierarchy of turbulent eddy will be dissipated by molecular viscosity in the dissipation range. Therefore, in the inertial subrange, all of the energy loss is energy dissipated inside the turbulent eddies during their breakage. We call this the energy dissipation of the inertial subrange, $W_{\mathrm{e\_\,inertia}}$.
  \item {\it Energy dissipation on the Kolmogorov scale (i.e., dissipation range)} $W_{\mathrm{e\_\,Kolmogorov}}$. As the breakage of turbulent eddies proceeds, their size will progressively decrease, so the viscosity effect becomes stronger. From a certain hierarchy of turbulent eddies, there is only the viscosity effect without the inertia effect. At this point, the turbulent eddies are located in the dissipation range. The size of this certain hierarchy of turbulent eddy is called the minimum size of the inertial subrange, and labeled as $\lambda_\mathrm{min}$ (Tennekes and Lumley, 1972). The minimum size of the dissipation range is indicated by the Kolmogorov scale $\eta$ (Pope, 2000), so the region $\left[\eta, \lambda_{\mathrm{min}}\right]$ is the dissipation range. In the dissipation range, all the kinetic energy of turbulent eddies will be transformed to internal energy by molecular viscosity, so we call this part the energy dissipation on the Kolmogorov scale (i.e., dissipation range) $W_{\mathrm{e\_\,Kolmogorov}}$. In fact, $W_{\mathrm{e\_\,Kolmogorov}}$ is the transfer energy from large turbulent eddies to smaller ones through their breakage in the inertial subrange.
  \item {\it Energy dissipation on the molecular scale} $W_\mathrm{\nu}$. The energy storage of energy-containing eddies $W_\mathrm{st}$, the energy dissipation of the inertial subrange $W_{\mathrm{e\_\,inertia}}$  and the energy dissipation on the Kolmogorov scale $W_{\mathrm{e\_\,Kolmogorov}}$ are the modes of energy action induced by the existence of turbulent eddies. Turbulence contains not only turbulent eddies but also a laminar component. The laminar component will also consume part of the energy in the system. The nature of this energy is identical to that of the energy consumption in fully laminar flow; namely, this energy is dissipated into internal energy by molecular viscosity and is not related to the structure of turbulent eddies. Therefore, we label the energy dissipated in the laminar component of turbulence by molecular viscosity as $W_\mathrm{\nu}$ and call it the energy dissipation on the molecular scale. Here, $\nu$ indicates typical viscosity dissipation, just like that in fully laminar flow.
\end{itemize}

So far, we have decomposed the energy of turbulence into different parts according to the characteristic size of turbulent eddies. In the following, how to quantify the different energy components of turbulence will be considered.
\subsubsection{Total energy of turbulence $W_\mathrm{T}$}

The total energy of turbulence in a volume of fluid per unit mass and per unit time $W_\mathrm{T}$ can be regarded as the sum of the energy storage of energy-containing eddies $W_\mathrm{st}$, the energy dissipation of the inertial subrange $W_\mathrm{e\_\,inertia}$, the energy dissipation on the Kolmogorov scale $W_\mathrm{e\_\,Kolmogorov}$ and the energy dissipation on the molecular scale $W_\mathrm{\nu}$,
\begin{equation}
  W_\mathrm{T}=W_\mathrm{\nu}+W_\mathrm{e\_\,Kolmogorov}+W_\mathrm{e\_\,inertia}+W_\mathrm{st}
\end{equation}

In a gas-fluid system, energy is mainly supplied by the expansion work of bubbles when they pass through the liquid layer. Based on this recognition, Bhavaraju et al. (1978) expressed the total energy consumption of gas-fluid systems as
\begin{equation}
  W_\mathrm{T}=\frac{\mathrm{ln}\left(P_1/P_2\right)}{P_1/P_2 -1}U_{\mathrm{g}}g
\end{equation}
Here, $P_1$ and $P_2$ are the pressure at the top and bottom of reactor, respectively, and $U_{\mathrm{g}}$ is the superficial velocity of bubbles. When Eq. (6) is used for a micro-control volume or in rough research, it can be assumed that $P_1=P_2$ . Then, the total energy consumption of a gas-fluid system can be rewritten as
\begin{equation}
   W_\mathrm{T}=U_{\mathrm{g}}g
\end{equation}

As mentioned above, the density of turbulent eddies is lower than that of the laminar component. Therefore, if we regard a turbulent eddy as a kind of ``bubble'', Eq. (7) can be used as an approximate measure of the total energy of turbulence, namely
\begin{equation}
   W_\mathrm{T}=U_{\mathrm{e}}g
\end{equation}
where $U_{\mathrm{e}}$ is the superficial velocity of turbulent eddies.
\subsubsection{Energy storage of energy-containing eddies $W_\mathrm{st}$}
In the range of energy-containing eddies, the surfaces of turbulent eddies are extremely unstable, which differs considerably from the situation of particles moving in gas but has some similarities to bubbles moving in liquid. In fact, there are some discrepancies between bubbles and turbulent eddies that move in liquid and the laminar regions of turbulence, respectively. Besides translational motion in the main direction of flow, there are some complicated secondary motions that occur in turbulent eddies, such as the violent oscillations of their surfaces. The energy contained in these secondary motions may be viewed as the energy storage of energy-containing eddies $W_\mathrm{st}$, which is also the total stored energy that can be transferred to smaller turbulent eddies.

Because the surface motion of turbulent eddies is extremely complicated, until now it has been impossible to directly compute the energy storage of energy-containing eddies. However, the resistant force $F_{\mathrm{D,eddy}}$  during the motion of turbulent eddies is composed of two parts, namely, the one caused by the body of turbulent eddies $F_{\mathrm{D,particle}}$ (we assume that it equals the drag force of solid particles of the same shape as the turbulent eddies), and that caused by the violent oscillations of the surfaces of turbulent eddies $F_{\mathrm{D,surf}}$. That is to say,
\begin{equation}
  F_{\mathrm{D,eddy}}=F_{\mathrm{D,particle}}+F_{\mathrm{D,surf}}
\end{equation}

When the eddies pass through laminar regions, the work of turbulence per unit mass can be expressed as
\begin{equation}
  W_{\mathrm{T}}=\frac{n_\mathrm{e}}{\left(1-f\right)\rho_{\mathrm{\,l}}+f\rho_{\mathrm{\,e}}}F_{\mathrm{D,eddy}}\cdot u_\mathrm{s}
\end{equation}
where $n_\mathrm{e}$ denotes the number of turbulent eddies of size $d_\mathrm{e}$ per unit volume. During this process, the energy stored in the violent oscillations of the surfaces of turbulent eddies can be expressed as
\begin{equation}
  W_{\mathrm{st}}=\frac{n_\mathrm{e}}{\left(1-f\right)\rho_{\mathrm{\,l}}+f\rho_{\mathrm{\,e}}}F_{\mathrm{D,surf}}\cdot u_\mathrm{s}
\end{equation}
If there is no surface oscillation, then there is no energy storage in the surface of turbulent eddies. Therefore, the energy storage in the surface of turbulent eddies should only be related to the force $F_{\mathrm{D,surf}}$ and the force $F_{\mathrm{D,particle}}$ can be excluded. Furthermore, from Eqs.(10)-(11) we obtain
\begin{equation}
  \frac{W_{\mathrm{st}}}{W_{\mathrm{T}}}=\frac{F_{\mathrm{D,surf}}}{F_{\mathrm{D,eddy}}}=\frac{F_{\mathrm{D,eddy}}-F_{\mathrm{D,particle}}}{F_{\mathrm{D,eddy}}}
\end{equation}
namely
\begin{equation}
  W_{\mathrm{st}}=W_{\mathrm{T}}\frac{F_{\mathrm{D,surf}}}{F_{\mathrm{D,eddy}}}=W_{\mathrm{T}}\frac{F_{\mathrm{D,eddy}}-F_{\mathrm{D,particle}}}{F_{\mathrm{D,eddy}}}
\end{equation}
Therefore, according to the discrepancy of drag force between solid particles and turbulent eddies (assuming that the solid particles have the same shape as the turbulent eddies), the energy storage of energy-containing eddies $W_{\mathrm{st}}$  can be approximated.

When turbulent eddies move in the laminar regions, their received resistant force can be approximated as
\begin{equation}
  F_{\mathrm{D,eddy}}=C_{\mathrm{D,eddy}}\frac{\pi}{4}d_{\mathrm{e}}^{2}\frac{\rho_{\mathrm{\,l}}}{2}u_{\mathrm{s}}^{2}
\end{equation}
Here, $C_{\mathrm{D,eddy}}$ is the drag coefficient of turbulent eddies and is assumed to equal that of bubbles (Yang et al., 2007). $F_{\mathrm{D,particle}}$ is the resistant force of solid particles that have the same shape as the turbulent eddies, and can be approximated as
\begin{equation}
  F_{\mathrm{D,particle}}=C_{\mathrm{D,particle}}\frac{\pi}{4}d_{\mathrm{e}}^{2}\frac{\rho_{\mathrm{\,l}}}{2}u_{\mathrm{s}}^{2}
\end{equation}
where $C_{\mathrm{D,particle}}$ is the drag coefficient of solid particles and can be calculated by the Schiller and Naumann (1935) equation:

\begin{equation}
{C_{\mathrm{D,particle}}} = \left\{ \begin{split}
&\frac{{24}}{{Re}} \left( {1 + 0.15R{e^{0.687}}} \right) &Re \le 1000\\
&0.44 & Re > 1000
\end{split} \right.
\end{equation}
Substituting Eqs. (14)-(15) into Eq. (13), the energy storage of energy-containing eddies can be finally expressed as
\begin{equation}
  W_{\mathrm{st}}=W_{\mathrm{T}} \frac{C_{\mathrm{D,eddy}}-C_{\mathrm{D,particle}}}{C_{\mathrm{D,eddy}}}
\end{equation}
\subsubsection{Energy dissipation of the inertial subrange $W_{\mathrm{e\_\,inertia}}$}
It is known that the energy dissipation of the inertial subrange $W_{\mathrm{e\_\,inertia}}$ mainly arises from the viscous dissipation occurring inside the turbulent eddies during their breakage. To quantify this energy consumption, we first have to develop a model for breakage of turbulent eddies, and determine the viscous dissipation inside the turbulent eddies during their breakage.

If we regard turbulent eddies as a kind of ``drop/bubble'', it is possible to use the current breakage model for a drop/bubble to represent the breakage of turbulent eddies (Lasheras et al., 2002; Liao and Lucas, 2009). We believe that the breakage of a turbulent eddy is caused by collision with another turbulent eddy that is equal to or smaller than the initial turbulent eddy and has sufficient kinetic energy. Turbulent eddies that are larger than the initial turbulent eddy merely transport it, similar to the assumption made for the breakage of a drop/bubble (Luo and Svendsen, 1996). Following the work of Luo and Svendsen (1996), the breakage model of a turbulent eddy with a diameter $d_\mathrm{e}$ can be expressed as follows:
\begin{equation}
  \Omega_{\mathrm{e}}\left(d_\mathrm{e}\right)=\int_{\lambda_{\mathrm{min}}}^{d_\mathrm{e}}\omega_{\mathrm{e,\lambda}}\left(d_\mathrm{e},\lambda\right)P_\mathrm{e}\left(d_\mathrm{e}|f_{\mathrm{BV}},\lambda\right)\mathrm{d}\lambda
\end{equation}
Here,  $\omega_{\mathrm{e,\lambda}}\left(d_\mathrm{e},\lambda\right)$ is the arrival (bombarding) frequency of turbulent eddies with a size between $\lambda$ and $\lambda+\mathrm{d}\lambda$ onto the initial turbulent eddy of size $d_\mathrm{e}$.  $P_\mathrm{e}\left(d_\mathrm{e}|f_{\mathrm{\,BV}},\lambda\right)$ is the probability of the initial turbulent eddy of size $d_\mathrm{e}$ breaking into two smaller turbulent eddies when the initial turbulent eddy is hit by an arriving turbulent eddy of size $\lambda$ that has a kinetic energy greater than or equal to the minimum energy required to induce breakage of the initial turbulent eddy. $f_{\mathrm{\,BV}}$ is the volume ratio of the smaller turbulent eddy to its mother turbulent eddy (i.e., the initial turbulent eddy of size $d_\mathrm{e}$)). Here, only binary breakage of the initial turbulent eddy is considered (Hagesaether et al., 2002).

Similar to gas kinetic theory, the collision frequency density can be determined as (Luo and Svendsen, 1996):
\begin{equation}
  \omega_{\mathrm{e,\lambda}}\left(d_\mathrm{e},\lambda\right)=\frac{\pi}{4}\left(d_\mathrm{e}+\lambda\right)^2{\bar u_\mathrm{\lambda}}{\dot n_\mathrm{\lambda}}n_\mathrm{e}
\end{equation}
where  ${\dot n_\mathrm{\lambda}}$ denotes the number of turbulent eddies with size between $\lambda$ and $\lambda+\mathrm{d}\lambda$ per unit volume, and  ${\bar u_\mathrm{\lambda}}$ is the turbulent velocity of turbulent eddies of size $\lambda$.

In the inertial subrange, the theory of isotropic turbulence can be used, so the mean turbulent velocity of turbulent eddies of size $\lambda$ is

\begin{equation}
  {\bar u_\mathrm{\lambda}}=\beta^{1/2}\left(\varepsilon \lambda\right)^{1/3}
\end{equation}
where $\beta =2$ and $\varepsilon$  is the dissipation rate of turbulent kinetic energy (Wang et al., 2003). The number density of turbulent eddies of size $\lambda$ is (Luo and Svendsen, 1996)
\begin{equation}
  {\dot n_\mathrm{\lambda}}=\frac{0.822\left(1-f\right)}{\lambda^4}
\end{equation}
and the number density of initial turbulent eddies of size $d_\mathrm{e}$ is (Vankova et al., 2007)
\begin{equation}
  n_\mathrm{e}=\frac{0.1\!\left(2\pi\right)^3}{d_{\mathrm{e}}^{4}}
\end{equation}
Substituting Eqs. (20)-(22) into Eq. (19), the collision frequency density is finally expressed as
\begin{equation}
  \omega_{\mathrm{e,\lambda}}\left(d_\mathrm{e},\lambda\right)=0.923\left(1-f\right)\frac{0.1\!\left(2\pi\right)^3}{d_{\mathrm{e}}^{4}}\varepsilon^{1/3}\frac{\left(d_\mathrm{e}+\lambda\right)^2}{\lambda^{11/3}}
\end{equation}
Most drop/bubble breakage models based on surface energy only take into account the energy constraint and predict a maximum breakage probability when $f_{\mathrm{\,BV}}$ approaches zero (Luo and Svendsen, 1996; Tsouris and Tavlarides, 1994). Wang et al. (2003) imposed both energy and capillary constraints. When a drop/bubble of size $d_{\mathrm{e}}$ is hit by a turbulent eddy of size $\lambda$ with kinetic energy of $e\left(\lambda\right)$, the daughter drop/bubble size has a minimum because of the capillary pressure and a maximum caused by the increase of surface energy. However, all of the above work neglects a part of energy consumption: the viscous dissipation inside a drop/bubble during its breakage (Vankova et al., 2007).
\begin{equation}
\begin{split}
  E_{\mathrm{DIS}}&=\frac{\pi}{6}d_{\mathrm{e}}^{3}\tau_\mathrm{D}=\frac{\pi}{6}d_{\mathrm{e}}^{3}\left[\left(\eta_\mathrm{D} \varepsilon^{1/3}d_{\mathrm{e}}^{1/3} \sqrt{\frac{\rho_{\mathrm{\,l}}}{\rho_{\mathrm{\,e}}}}\;\right)/d_\mathrm{e}\right] \\
  &=\frac{\pi}{6}\eta_\mathrm{D} \varepsilon^{1/3} d_{\mathrm{e}}^{7/3} \sqrt{\frac{\rho_{\mathrm{\,l}}}{\rho_{\mathrm{\,e}}}}
\end{split}
\end{equation}
Here, $\tau_\mathrm{D}$ is the viscous stress inside the breaking drop/bubble, which is estimated as proposed by Davies (1985), and $\eta_\mathrm{D}$ is the viscosity of the drop/bubble.

Because we regard turbulent eddies as a kind of drop/bubble, when a turbulent eddy collides with another turbulent eddy of equal or smaller size, the condition for the initial oscillating deformed turbulent eddy to break is that the kinetic energy of the bombarding turbulent eddy $e\left(\lambda\right)$ of size $\lambda$ exceeds the increase in surface energy required for breakage. Meanwhile, the dynamic pressure $0.5\rho_\mathrm{e}{\bar u_{\lambda}^{\,2}}$  must exceed the capillary pressure $\sigma /r$, where $\sigma$ denotes the interfacial tension of turbulent eddies and $r$ is their radius of curvature. Moreover, the kinetic energy of bombarding turbulent eddies must also provide the energy dissipated inside the initial turbulent eddy during its breakage. That is to say,
\begin{equation}
  e\left(\lambda\right)\ge \mathrm{max}\left(c_{\!f_{\mathrm{\,BV}}}\pi d_{\mathrm{e}}^{\,2}\sigma , \frac{\pi \sigma \lambda^3}{3d_{\mathrm{e}} {f_{\mathrm{\,BV}}}^{1/3}}\right)+E_{\mathrm{DIS}}
\end{equation}

It should be noted that this breakage mechanism of turbulent eddies involves breakage caused by collision with eddies of equal or smaller size, which does not mean that the energy of the smaller turbulent eddies will be transferred to the larger ones. If this occurred, it would contradict the picture of the turbulent eddy cascade, which is that energy is transferred from the larger turbulent eddies to smaller ones. In fact, the increase in surface energy required for the breakage of large turbulent eddies is still transferred from the larger turbulent eddies, which is the first part of energy loss for each hierarchy. Finally, the transfer energy enters the dissipation range and becomes the kinetic energy of turbulent eddies.

To determine the energy contained in turbulent eddies of different size, a distribution function of kinetic energy for turbulent eddies is required. The normalized exponential energy density function (Hagesaether et al., 2002) is regarded as a suitable model for the kinetic energy of turbulent eddies, namely
\begin{equation}
  p_\mathrm{e}\left(\chi \right)=\mathrm{exp}\left(-\chi\right)
\end{equation}
where  $\chi=\frac{e\left(\lambda\right)}{{\bar e\left(\lambda\right)}}$. Consequently, the probability density function for a turbulent eddy of size $d_\mathrm{e}$ to break with a breakage fraction $f_{\mathrm{\,BV}}$ when it is hit by another turbulent eddy of equal or smaller size $\lambda$ can be expressed as follows
\begin{equation}
  P_{\mathrm{\tilde e}}\left(d_\mathrm{e}|f_{\mathrm{\,BV}},\lambda\right)=p_\mathrm{e}\left[e\left(\lambda\right)\ge\mathrm{max}\left(c_{\!f_{\mathrm{\,BV}}}\pi d_{\mathrm{e}}^{\,2}\sigma , \frac{\pi \sigma \lambda^3}{3d_{\mathrm{e}} {f_{\mathrm{\,BV}}}^{1/3}}\right)+E_{\mathrm{DIS}}\right]
\end{equation}
Furthermore, according to probability theory, the probability of this current turbulent eddy breaking is
\begin{equation}
  P_{\mathrm{e}}\left(d_\mathrm{e}|f_{\mathrm{\,BV}},\lambda\right)=\int_{0}^{0.5}P_{\mathrm{\tilde e}}\left(d_\mathrm{e}|f_{\mathrm{\,BV}},\lambda\right)\mathrm{d}f_{\mathrm{\,BV}}
\end{equation}
Substituting Eqs. (23) and (28) into Eq. (18), the breakage model of a turbulent eddy of size $d_\mathrm{e}$ is obtained as follows
\begin{equation}
  \Omega_\mathrm{e}\left(d_\mathrm{e}\right)=\int_{\lambda_{\mathrm{min}}}^{d_\mathrm{e}}\!\int_{0}^{0.5}\!\!\!\omega_{\mathrm{e,\lambda}}\left(d_\mathrm{e},\lambda\right)P_\mathrm{e}\left(d_\mathrm{e}|f_{\mathrm{\,BV}},\lambda\right)\mathrm{d}f_{\mathrm{\,BV}}\mathrm{d}\lambda
\end{equation}

Therefore, in a volume of turbulent fluid per unit mass, the viscous dissipation occurring inside a turbulent eddy during its breakage is finally expressed as
\begin{equation}
W_{\mathrm{e\_\,inertia}}=\!\!\int_{\lambda_{\mathrm{min}}}^{d_\mathrm{e}}\!\int_{0}^{0.5}\!\!\!\!\!\frac{1}{\left(1-f\right)\rho_\mathrm{l}+f\rho_\mathrm{e}}\omega_{\mathrm{e,\lambda}}\left(d_\mathrm{e},\lambda\right)P_\mathrm{e}\left(d_\mathrm{e}|f_{\mathrm{\,BV}},\lambda\right)E_{\mathrm{DIS}}\mathrm{d}f_{\mathrm{\,BV}}\mathrm{d}\lambda
\end{equation}
\subsubsection{Energy dissipation on the Kolmogorov scale $W_{\mathrm{e\_\,Kolmogorov}}$}

The mean velocity of turbulent eddies of size $\lambda$ in the inertial subrange of isotropic turbulence has been expressed by Eq. (20). We assume that this expression is also valid in the dissipation range of isotropic turbulence. Therefore, the mean kinetic energy ${\bar e\left(\lambda\right)}$  of a turbulent eddy of size $\lambda$ in the dissipation range can be given as
\begin{equation}
  {\bar e}\left(\lambda\right)=\rho_{\mathrm{\,l}}\frac{\pi}{6}\lambda^3\frac{{\bar u}_{\mathrm{\lambda}}^{2}}{2}=\frac{\pi \beta}{12}\rho_{\mathrm{\,l}}\varepsilon^{2/3}\lambda^{11/3}
\end{equation}

There is no breakage of turbulent eddies in the dissipation range, so all of the energy contained in these turbulent eddies will be transferred into internal energy by molecular viscosity. That is, the energy dissipation on the Kolmogorov scale $W_{\mathrm{e\_
\,Kolmogorov}}$ can be regarded as the sum of the kinetic energy of turbulent eddies in the dissipation range. This kind of energy is instantaneously and completely transferred into internal energy. Meanwhile, to maintain the existence of the dissipation range, a new compensatory kinetic energy supplying the turbulent eddies in this subrange is received continuously from the inertial subrange through the breakage of turbulent eddies. Therefore,
\begin{equation}
  W_{\mathrm{e\_\,Kolmogorov}}=\int_{k_{\mathrm{\lambda_{min}}}}^{k_\mathrm{\eta}}\!\!\!{\bar e}\left(\lambda\right)\mathrm{d}n_\mathrm{e}
\end{equation}
where $\mathrm{d}n_\mathrm{e}$ denotes the number density of turbulent eddies with size between $d_\mathrm{e}$ and $d_\mathrm{e}+\mathrm{d}d_\mathrm{e}$ in the dissipation range, and $k_\mathrm{\eta}$ and $k_{\mathrm{\lambda_{min}}}$ are the wave numbers corresponding to size $\eta$ and $\lambda_{\mathrm{min}}$ respectively.

The minimum size of turbulent eddies $\lambda_{\mathrm{min}}$ in the inertial subrange should be taken as the upper limit of the dissipation range. Here, $\lambda_{\mathrm{min}}=60\eta$ is assumed because the range $\left[\eta,60\eta\right]$ is regarded as the dissipation range. A way to obtain a more precise value of $\lambda_{\mathrm{min}}$ still needs to be developed. Meanwhile, the number density of turbulent eddies $\mathrm{d}n_\mathrm{e}$ within a given size range can be obtained by integrating the energy spectrum, namely (Vankova et al., 2007)
\begin{equation}
  \frac{\mathrm{d}n_\mathrm{e}}{\mathrm{d}k}=0.1k^2
\end{equation}
where $k$ is the wave number and equals $2\pi/\lambda$.

Based on Eqs. (31)-(33), in a volume of turbulent fluid per unit mass, the energy dissipation on the Kolmogorov scale $W_{\mathrm{e\_\,Kolmogorov}}$ can be finally expressed as
\begin{equation}
  W_{\mathrm{e\_\,Kolmogorov}}=\int_{\eta}^{\lambda_{\mathrm{min}}}\!\frac{\pi\beta}{12}\rho_\mathrm{\,l}\,\varepsilon^{2/3}\lambda^{11/3}\,0.1\frac{\left(2\pi\right)^3}{\lambda^4}\,\frac{f}{\left(1-f\right)\rho_{\mathrm{\,l}}+f\rho_{\mathrm{\,e}}}\mathrm{d}\lambda
\end{equation}

\begin{figure*}[htbp]
  \centering
  \includegraphics[width=0.85\textwidth]{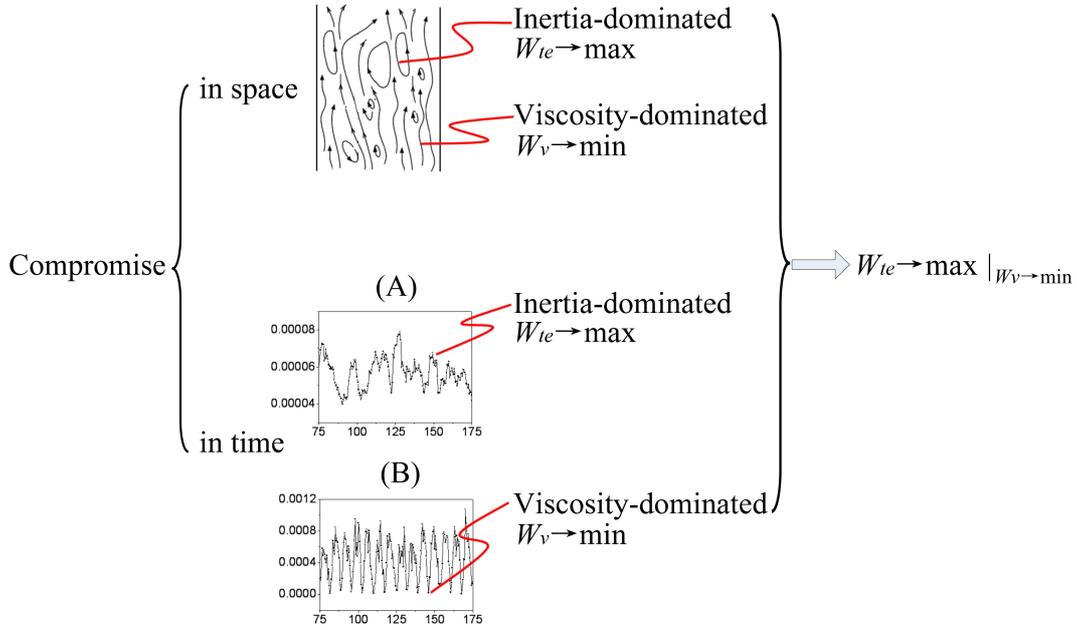}
  \caption{Compromise between inertial and viscosity effects of turbulence in space and time.}\label{fig3}
\end{figure*}

\subsection{A stability condition for turbulence}

When the turbulence in a pipe is considered, the boundary layer near the pipe wall is dominated by viscosity, while the internal region shows inviscid characteristics because inertia dominates. When viscosity dominates, the principle of least viscous dissipation (Lamb, 1932) or minimum entropy production (Prigogine, 1945) can be applied. Then, our question is: when the effect of inertia dominates, what kind of mechanism should be used? Li et al. (1999b) assumed a tendency of maximum total dissipation $W_\mathrm{T}$ in turbulent pipe flow. They also pointed out that for all real cases of turbulence in a pipe, neither viscosity nor inertia could exclusively dominate the system; they had to compromise with each other in realizing their respective intrinsic tendencies. This is the so-called turbulence stability condition, which is based on an extremum control mechanism. Finally, by assuming a general expression for both laminar and turbulent flow in a pipe, they calculated the radial velocity profiles of water and air in a pipe at different $Re$ based on the turbulence stability condition. They compared the results with experimental data obtained by von Karman (1939) and Bejan (1982), which demonstrated that the turbulence stability condition allowed reasonable analysis of the turbulence in a pipe.

Wang et al. (2007; Wang, 2008) extended the work of Li et al. (1999b) and proposed that the compromise between viscosity and inertia might be of general relevance to fluid flow. They thought that the dissipation associated with a time-averaged flow field might characterize the viscosity effect, while the dissipation associated with the temporal variation of velocity (i.e., fluctuation) might characterize the inertia effect. It has been already reported that on the micro-scale, flow could be either viscosity-dominated inside a turbulent eddy or inertia-dominated at the interface between turbulent eddies (Li and Kwauk, 2003), while on the macro-scale, flow was dominated by inertia in the core region, and viscosity in the wall region. Therefore, Wang et al. (2007; Wang, 2008) used the maximization of turbulent dissipation $W_{\mathrm{te}}$ as a more general and appropriate index of the effect of inertia on turbulent flow than $W_\mathrm{T}$; that is, they extended the turbulence stability condition. Finally, to verify their improved turbulence stability condition, Wang et al. performed DNS of the flow around a cylinder using a macro-scale particle method (Ma et al., 2006). Statistical data for viscous and turbulent dissipation validated the correctness of their work (Fig. 3).

Concerning the separation of energy of turbulence into different components, the energy dissipation of the inertia effect includes the energy dissipation of the inertial subrange $W_{\mathrm{e\_\,inertia}}$ and the energy dissipation on the Kolmogorov scale $W_{\mathrm{e\_\,Kolmogorov}}$, namely
\begin{equation}
  W_{\mathrm{te}}=W_{\mathrm{e\_\,Kolmogorov}}+W_{\mathrm{e\_\,inertia}}
\end{equation}
In contrast, the energy storage of energy-containing eddies $W_{\mathrm{st}}$ is not transformed into internal energy at the current time, but is later transferred to smaller turbulent eddies, so it does not belong to the energy dissipation of the inertia effect $W_{\mathrm{te}}$. The dissipation of the viscosity effect in the turbulence stability condition is just the energy dissipation on the molecular scale $W_\mathrm{\nu}$ for the laminar component. Then, under the current energy decomposition framework, the turbulence stability condition of Li et al. (1999b; Wang et al., 2007; Wang, 2008) can be expressed as
\begin{equation}
  W_{\mathrm{\nu}}\to\mathrm{min}|_{\,W_\mathrm{te}\to\mathrm{max}}
\end{equation}

\begin{figure}[htbp]
  \centering
  \includegraphics[width=0.85\textwidth]{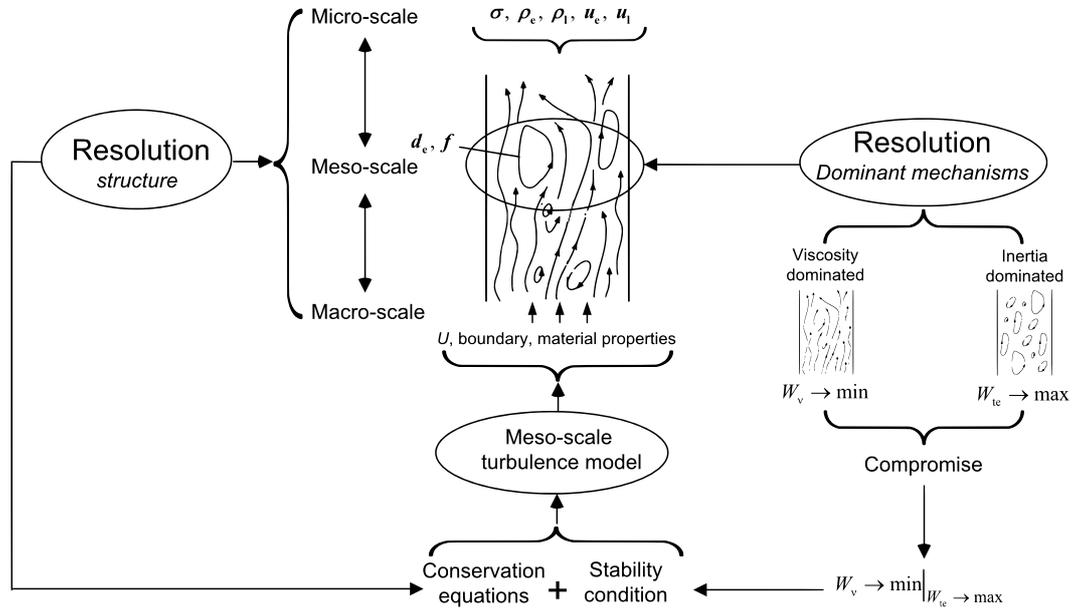}
  \caption{The physical basis of the EMMS-based turbulence model.}\label{fig4}
\end{figure}

\begin{table}
\centering
\caption{Summary of the formulae of the EMMS-based turbulence model}
\begin{footnotesize}
  \begin{tabular}{l}
    \toprule
    Constraint equation of turbulent eddies:\\
    $\frac{\pi}{6}d_{\mathrm{e}}^{3}\left(\rho\,_\mathrm{l}-\rho\,_\mathrm{e}\right)g=C_\mathrm{D,eddy}\frac{\pi}{4}d_{\mathrm{e}}^{3}\frac{\rho\,_\mathrm{l}}{2}u_\mathrm{s}^2$\\
     Total energy of turbulence:\\
    $W_\mathrm{T}=U_{\mathrm{e}}g$\\
   Energy decomposition of turbulence:\\
    $W_{\mathrm{T}}=W_{\mathrm{\nu}}+W_{\mathrm{e\_\,innertia}}+W_{\mathrm{e\_\,Kolmogorov}}+W_{\mathrm{st}}$\\
   Energy dissipation in the inertial subrange:\\
   $W_{\mathrm{e\_\,inertia}}\!\!=\!\!\int_{\lambda_{\mathrm{min}}}^{d_\mathrm{e}}\!\!\int_{0}^{0.5}\!\!\!\!\!\frac{1}{\left(1-f\right)\rho_\mathrm{l}+f\rho_\mathrm{e}}\omega_{\mathrm{e,\lambda}}\left(d_\mathrm{e},\lambda\right)P_\mathrm{e}\left(d_\mathrm{e}|f_{\mathrm{\,BV}},\lambda\right)E_{\mathrm{DIS}}\mathrm{d}f_{\mathrm{\,BV}}\mathrm{d}\lambda$\\
    Energy dissipation on the Kolmogorov scale:\\
    $W_{\mathrm{e\_\,Kolmogorov}}\!=\!\!\int_{\eta}^{\lambda_{\mathrm{min}}}\!\!\frac{\pi\beta}{12}\rho_\mathrm{\,l}\,\varepsilon^{2/3}\lambda^{11/3}\,0.1\frac{\left(2\pi\right)^3}{\lambda^4}\frac{f}{\left(1-f\right)\rho_{\mathrm{\,l}}+f\rho_{\mathrm{\,e}}}\mathrm{d}\lambda$\\
    Turbulence stability condition:\\
    $W_{\mathrm{\nu}}\to\mathrm{min}|_{\,W_\mathrm{te}\to\mathrm{max}}$\\
    \bottomrule
  \end{tabular}
\end{footnotesize}
\end{table}
%
%
\begin{table}
  \centering
  \caption{Summary of other expressions used to close the EMMS-based turbulence model}
  \begin{footnotesize}
    \begin{tabular}{m{0.3\textwidth}m{0.1\textwidth}m{0.3\textwidth}}
  \toprule
    $\rho_{\mathrm{\,e}}=0.99\rho_{\mathrm{\,l}}$ && $u_\mathrm{s}=\frac{U_\mathrm{e}}{f}-\frac{U_\mathrm{l}}{1-f}$\\
    $W_{\mathrm{st}}=W_\mathrm{T}\frac{C_{\mathrm{D,eddy}}-C_{\mathrm{D,particle}}}{C_{\mathrm{D,eddy}}}$ & & $\eta=\frac{\left(\mu_{\mathrm{\,l}}/\rho_{\mathrm{\,l}}\right)^{3/4}}{\varepsilon^{1/4}}$\\
     $\varepsilon=W_\mathrm{T}-W_\mathrm{\nu}$
    & & $\lambda_{\mathrm{min}}=60\eta$\\
    \multicolumn{2}{l}{$\omega_{\mathrm{e,\lambda}}\left(d_\mathrm{e},\lambda\right)=0.923\left(1-f\right)\frac{0.1\!\left(2\pi\right)^3}{d_{\mathrm{e}}^{4}}\varepsilon^{1/3}\frac{\left(d_\mathrm{e}+\lambda\right)^2}{\lambda^{11/3}}$}\\
    \multicolumn{2}{l}{$P_{\mathrm{\tilde e}}\left(d_\mathrm{e}|f_{\mathrm{\,BV}},\lambda\right)=p_\mathrm{\,e}\left[e\left(\lambda\right)\ge\mathrm{max}\left(c_{_{f_{\mathrm{\,_{BV}}}}}\pi d_{\mathrm{e}}^{\,2}\sigma , \frac{\pi \sigma \lambda^3}{3d_{\mathrm{e}} {f_{\mathrm{\,BV}}}^{1/3}}\right)+E_{\mathrm{DIS}}\right]$}\\
    $E_{\mathrm{DIS}}=\frac{\pi}{6}\eta_{_\mathrm{D}}\varepsilon^{1/3}d_{\mathrm{e}}^{\,7/3}\sqrt{\frac{\rho_{\mathrm{\,l}}}{\rho_{\mathrm{\,e}}}}$
    & & $c_{_{f_{\mathrm{\,_{BV}}}}}=f_{\mathrm{BV}}^{\,2/3}+\left(1-f_{\mathrm{\,BV}}\right)^{2/3}-1$\\
    \multicolumn{2}{l}{\rule{0pt}{5ex} $ C_{\mathrm{D,\,particle}} =\left\{\begin{aligned}
    &24/Re  \left( {1 + 0.15R{e^{0.687}}} \right)  &Re \le 1000\\
    &0.44  &Re > 1000
     \end{aligned}\right.$}\\
    $C_{\mathrm{D,\,eddy}}=C_{\mathrm{D0,\,eddy}}\left(1-f\right)^{p}$
    && $C_{\mathrm{D0,\,eddy}}=\frac{4}{3}\frac{g d_\mathrm{e}}{U_{\mathrm{T}}^{2}}\frac{\rho_{\mathrm{\,l}}-\rho_{\mathrm{\,e}}}{\rho_{\mathrm{\,l}}}$\\
  \multicolumn{2}{l}{\rule{0pt}{5ex} $ J = \left\{ \begin{aligned}
   &0.94{H^{\,0.757}} & 2 < H \le 59.3\\
   &3.42{H^{\,0.441}} & H > 59.3
\end{aligned} \right.$}\\
  $U_\mathrm{T}=\frac{\mu_{\mathrm{\,l}}}{\rho_{\mathrm{\,l}}\, d_{\mathrm{e}}}M_o^{-0.149}\left(J-0.857\right)$
  & & $H=\frac{4}{3}E_\mathrm{o} M_\mathrm{o}^{\,-\,0.149} \left(\frac{\mu_{\mathrm{\,l}}}{0.0009}\right)^{-0.14}$\\
  $E_\mathrm{o}=\frac{d_{\mathrm{e}}^{2} \left(\rho_{\,\mathrm{l}}-\rho_{\,\mathrm{e}}\right) g}{\sigma}$
  & & $M_\mathrm{o}=\frac{g \mu_{\,\mathrm{l}}^{4}\left(\rho_{\,\mathrm{l}}-\rho_{\,\mathrm{e}}\right)}{\rho_{\,\mathrm{l}}^{2}\sigma^3}$\\
  \bottomrule
  \end{tabular}
  \end{footnotesize}
\end{table}

\subsection{Physical basis and summary of the EMMS-based turbulence model}

As depicted schematically in Fig. 4, the physical basis of the EMMS-based turbulence model is described by multi-scale analysis, which resolves the system into three basic scales: molecular scale, eddy scale and vessel scale; that is, the micro-, meso- and macro-scales, respectively. This physical description allows the overall concept of the EMMS-based turbulence model to be easily understood, as well as the logical relationships among different specific parts such as the inhomogeneous structural parameters of turbulence, the constraint equations of turbulent eddies, and the turbulence stability condition. Specifically, we first decompose the flow system from the aspects of scale and control mechanism, and then describe different control mechanisms as a corresponding extremum tendency where the compromise between these tendencies forms the stability condition of the system. Mathematically, this formulation can be expressed as a multi-objective variational problem in which each control mechanism is a conditional extremum subject to the other control mechanisms, so dynamic constraint equations at different scales are related to form a closed model. In Fig. 4, the turbulence stability condition plus five conservation constraint equations are used to produce the variational criterion, leading to the EMMS-based turbulence model. Table 1 summarizes the formulae of the EMMS-based turbulence model, which is a mathematical description for the physical basis of the model. Other relevant expressions used to close the EMMS-based turbulence model are summarized in Table 2.

\begin{figure}[htbp]
  \centering
  \includegraphics[width=0.6\textwidth]{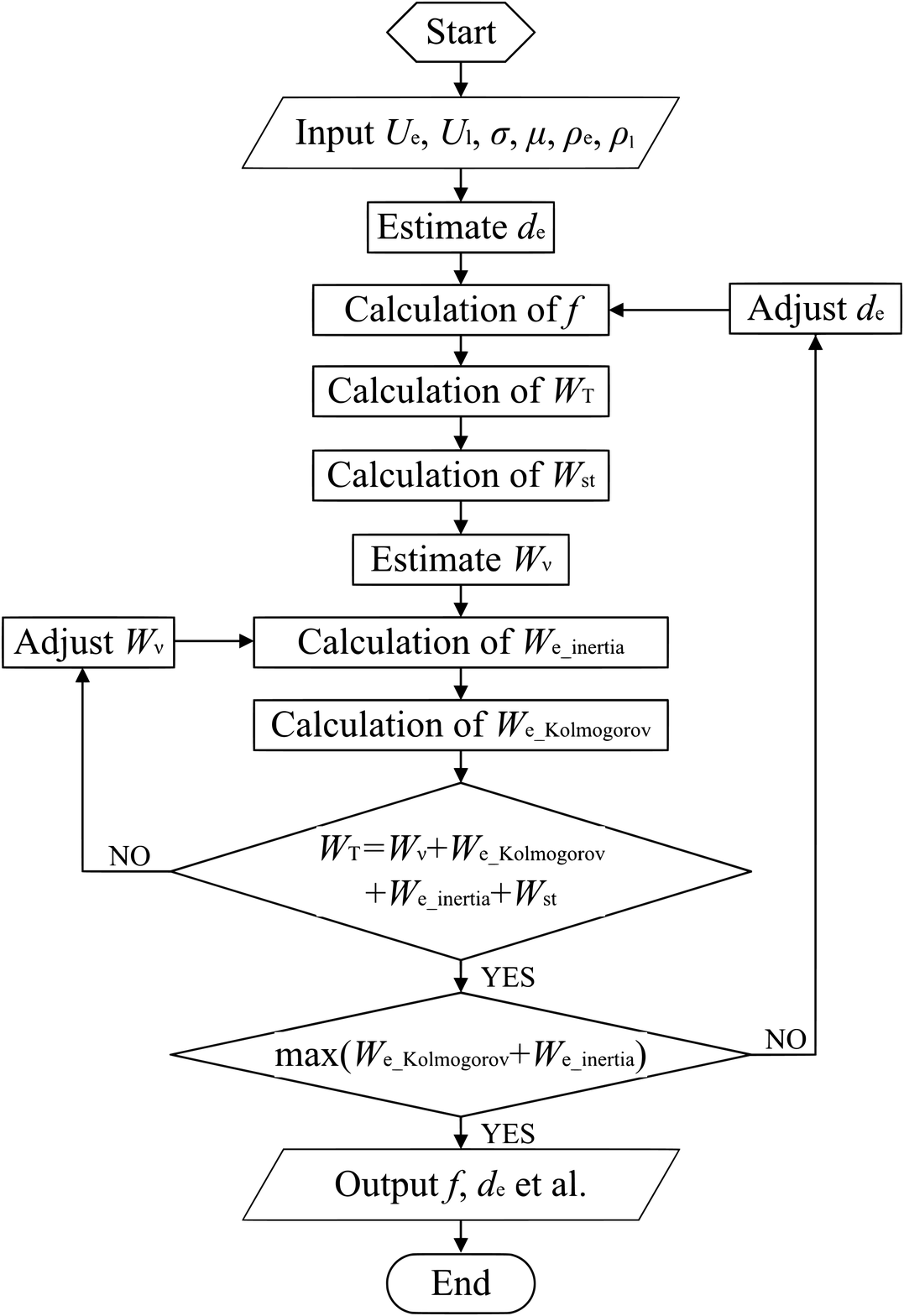}
  \caption{Flow chart describing solution of the EMMS-based turbulence model.}\label{fig5}
\end{figure}

\subsection{Flow chart describing the EMMS-based turbulence model}

A flow chart outlining the EMMS-based turbulence model is presented in Fig. 5. For a given flow system, the inhomogeneous structural parameters of the flow state can be computed in advance using the EMMS-based turbulence model with the traverse method. A mesh table is then established, which is convenient to incorporate into computational fluid dynamics (CFD). This allows two inhomogeneous structural parameters, $d_\mathrm{e}$ and $f$, to be optimized. The flow chart can be summarized as follows:
\begin{enumerate}[\ \ 1)]
  \item Input initial data such as the superficial velocity of turbulent eddies $U_{\mathrm{e}}$, the superficial velocity of the laminar component $U_{\mathrm{l}}$, the interfacial tension of turbulent eddies $\sigma$, the density of turbulent eddies $\rho_{\mathrm{e}}$, and the density of the laminar component $\rho_{\mathrm{l}}$.
  \item Estimate a value for the equivalent diameter of turbulent eddies $d_{\mathrm{e}}$.
  \item Calculate the volume fraction of turbulent eddies $f$ using Eq. (1), the total energy of turbulence $W_{\mathrm{T}}$ using Eq. (8), and the energy storage of energy-containing eddies $W_{\mathrm{st}}$ using Eq. (17).
  \item Estimate a value for the energy dissipation on the molecular scale $W_{\mathrm{\nu}}$.
  \item Calculate the energy dissipation of the inertial subrange $W_{\mathrm{e\_\,inertia}}$ using Eq. (30), and the energy dissipation on the Kolmogorov scale $W_{\mathrm{e\_\,Kolmogorov}}$ using Eq. (34).
  \item If the total energy of turbulence $W_{\mathrm{T}}$ computed by Eq. (8) is equal to the sum of the energy dissipation on the molecular scale $W_{\mathrm{\nu}}$, the energy dissipation on the Kolmogorov scale $W_{\mathrm{e\_\,Kolmogorov}}$, the energy dissipation of the inertial subrange $W_{\mathrm{e\_\,inertia}}$ and the energy storage of energy-containing eddies $W_{\mathrm{st}}$, namely
      \begin{equation*}
        W_{\mathrm{T}}=W_{\mathrm{\nu}}+W_{\mathrm{e\_\,Kolmogorov}}+W_{\mathrm{e\_\,inertia}}+W_{\mathrm{st}},
      \end{equation*}
      then go to step 7); otherwise, adjust the value of $W_{\mathrm{\nu}}$ and go to step 5).
  \item	If the sum of $W_{\mathrm{e\_\,Kolmogorov}}$ and $W_{\mathrm{e\_\,inertia}}$ is its maximum value, so
      \begin{equation*}
        \max\left(W_{\mathrm{e\_\,Kolmogorov}}+W_{\mathrm{e\_\,inertia}}\right)
      \end{equation*}
      is true, then go to step 8); otherwise, adjust the value of $d_{\mathrm{e}}$ and go to step 3).
  \item Output the optimized value for the volume fraction of turbulent eddies $f$ and that for the equivalent diameter of turbulent eddies $d_{\mathrm{e}}$.
\end{enumerate}

\section{Results and discussions}
\label{}
\subsection{ Results for the EMMS-based turbulence model}
\label{}

In this paper, the superficial velocity of the laminar component $U_{\mathrm{l}}$ was fixed at 0.001 m/s, while the superficial velocity of turbulent eddies $U_{\mathrm{e}}$ was varied from 0.01 to 3.0 m/s. Here, we took a turbulent jet emerging from an orifice into a tank for example and thought that the still fluid in the tank was the non-turbulent eddies, so $U_{\mathrm{l}}$  was chosen as a small value near 0. Meanwhile, the moving fluid emerging from the orifice was regarded as the turbulent eddies. The density of the laminar component was $\rho_{\mathrm{l}}$ = 1000 kg/$\mathrm{m^3}$, and the density of turbulent eddies was ${\rho_{\mathrm{\,e}}} = 0.99{\rho_{\mathrm{\,l}}} = 990 \ \mathrm{kg/{m^3}}$. The surface tension of turbulent eddies was fixed at that of a water drop, namely $\sigma$=0.075 N/m, and the viscosity coefficient of the laminar component was $\mu  = 1.00374 \times {10^{ - 3}}$ $\mathrm{Pa \cdot s}$, which was also used for the turbulent eddies.

Fig. 6 presents the results obtained for the inhomogeneous structural parameters $f$ and $d_{\rm{e}}$ for different inlet velocity ${U_{\rm{in}}}$ optimized by the EMMS-based turbulence model. Fig. 6(a) reveals that as ${U_{\rm{in}}}$ increases, the volume fraction of turbulent eddies $f$ increases correspondingly. When ${U_{\rm{in}}}$ is small, the flow state is laminar almost everywhere, so the volume fraction of turbulent eddies $f$ increases rapidly with increasing ${U_{\rm{in}}}$. When ${U_{\rm{in}}}$ is large, the flow state is turbulent eddies almost everywhere, so the volume fraction of turbulent eddies $f$ increases only slightly as ${U_{\rm{in}}}$ increases further and $f$ finally approaches 1. Fig. 6(b) indicates that the equivalent diameter of turbulent eddies $d_{\rm{e}}$ first decreases with increasing ${U_{\rm{in}}}$, and then begins to increase slightly. When ${U_{\rm{in}}}$ is small, the flow state is laminar almost everywhere, so a small increase of ${U_{\rm{in}}}$ will readily decrease the minimum size of turbulent eddies (i.e., the Kolmogorov scale $\eta$), which leads to the decrease of the equivalent diameter of turbulent eddies $d_{\rm{e}}$. When the minimum size of turbulent eddies is reached, the dissipation rate of turbulence becomes large, which leads to extra energy transfer in the inertial subrange and forces more energy to be transferred into the dissipation range. Based on this understanding, the equivalent diameter of turbulent eddies $d_{\rm{e}}$ will increase slightly as ${U_{\rm{in}}}$ increases (see the right side of the lowest peak). In fact, this slight increase is only for the equivalent diameter of turbulent eddies $d_{\rm{e}}$, and the minimum size of turbulent eddies always decreases a little as ${U_{\rm{in}}}$ increases.

\subsection{Numerical examples improved by the EMMS-based turbulence model}
To validate the EMMS-based turbulence model, two numerical examples, namely, lid-driven cavity flow and turbulent flow with forced convection in an empty room, were simulated and the calculated results compared with experimental data.

In general, the effective kinematic viscosity ${\nu _{\rm{eff}}}$ is expressed as
\begin{equation}
  {\nu _{\rm{eff}}} = {\nu _{\rm{0}}} + {\nu _{\rm{t}}}
\end{equation}
where ${\nu _{\rm{0}}}$ is the kinematic viscosity of a fluid and ${\nu _{\rm{t}}}$ is turbulent kinematic viscosity.

\begin{figure}[tpb]
  \centering
  \includegraphics[width=0.5\textwidth]{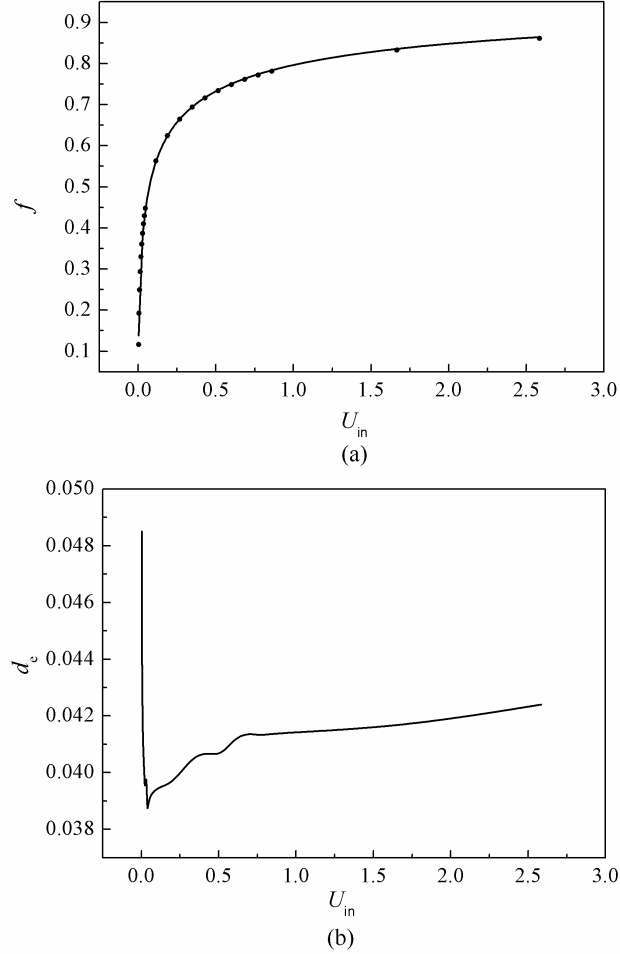}
  \caption{The results of EMMS-based turbulence model}
\end{figure}

To make the effective kinematic viscosity ${\nu _{\rm{eff}}}$ contain the information of inhomogeneous structural parameters, we rewrote it according to the two-phase concept of turbulence:
\begin{equation}
\nu_{\rm{eff}} = \left( {1 - f} \right)\nu_{\rm{0}} + f\nu_{\rm{t}}
\end{equation}
including the inhomogeneous structural parameter $f$. Theoretically, traditional turbulence models can be improved by incorporating those turbulent structural parameters. In the following, a zero-equation model(Chen and Xu, 1998) and the standard $k$-$\varepsilon$ model (Launder and Spalding, 1974) were improved to simulate lid-driven cavity flow and turbulent flow with forced convection in an empty room, respectively. The EMMS-based turbulence model is referred to the turbulence models improved by incorporating turbulent structural parameters and its ${\nu _{\rm{eff}}}$ is calculated by Eq.(38).

The pressure-implicit with splitting of operators (PISO) algorithm, which is part of the open source CFD software package OpenFOAM (OpenCFD and Ltd, 2009), was used to solve the Navier-Stokes equations.

\subsubsection{Lid-driven cavity flow}
Lid-driven cavity flow is a classical benchmark problem for the evaluation of numerical methods (Botella and Peyret, 1998). Therefore, we simulated a two-dimensional lid-driven cavity flow incorporating the EMMS-based turbulence model. First, a zero-equation model (Chen and Xu, 1998) was revised. The benchmark data obtained by Ghia et al. (1982) was used as the reference solution.

Fig. 7 shows the simulated geometry of lid-driven cavity flow, in which the upper wall moves at a constant velocity of $U=1.0$ m/s toward positive $x$ direction and the other walls are fixed with no-slip boundary conditions.
\begin{figure}[htbp]
  \centering
  \includegraphics[width=0.6\textwidth]{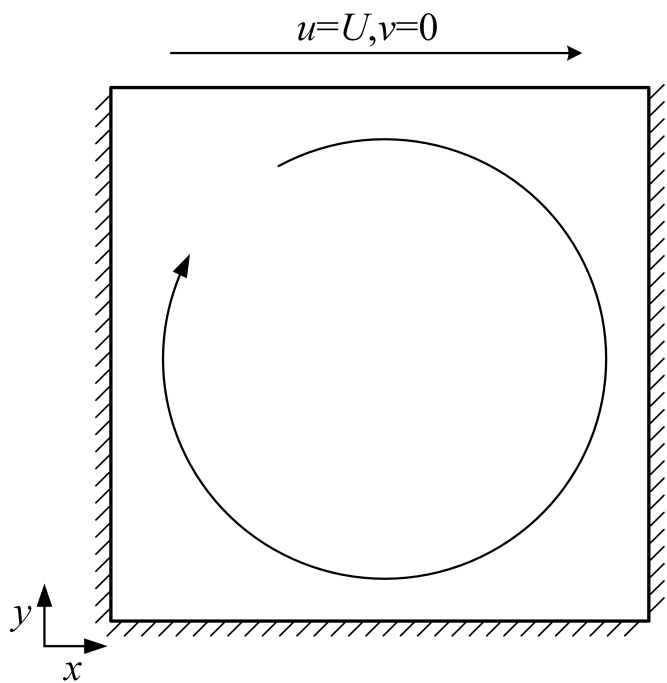}
  \caption{Diagram of lid-driven cavity flow.}\label{fig7}
\end{figure}

The computed velocity is compared with the reference solution at the sections of $x$=\,0.5 and $y$=\,0.5 for Reynolds numbers $Re$ of 1000, 5000, 7500 and 10000 in Fig. 8. Our simulated results agree well with the reference solution at relatively low Reynolds number ($Re$$\le$\,5000). Even for relatively large Reynolds number such as $Re$=7500 and 10000, our computed results are in reasonable agreement with the reference solution; some discrepancy emerges especially at the section of $y$=\,0.5 near the position of $x$=\,0.05 (see Fig. 8(b)).

\begin{figure}[htbp]
  \centering
  \includegraphics[width=0.8\textwidth]{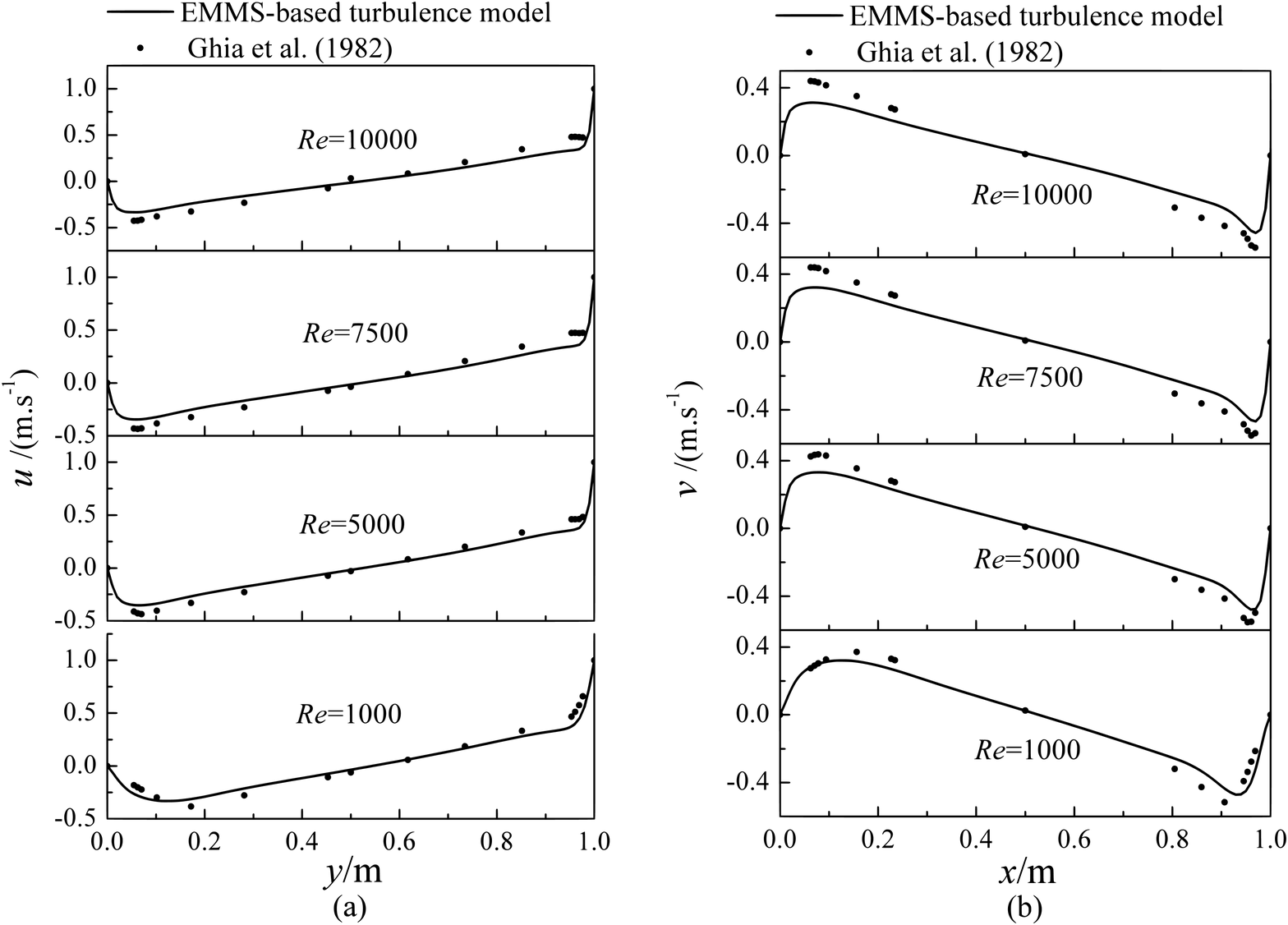}
  \caption{Comparison of computed velocity at different Reynolds number with the benchmark data of Ghia et al. (1982) at sections of (a) $x=0.5$, and (b) $y=0.5$.}\label{fig8}
\end{figure}

\begin{figure}[htbp]
  \centering
    \includegraphics[width=0.9\textwidth]{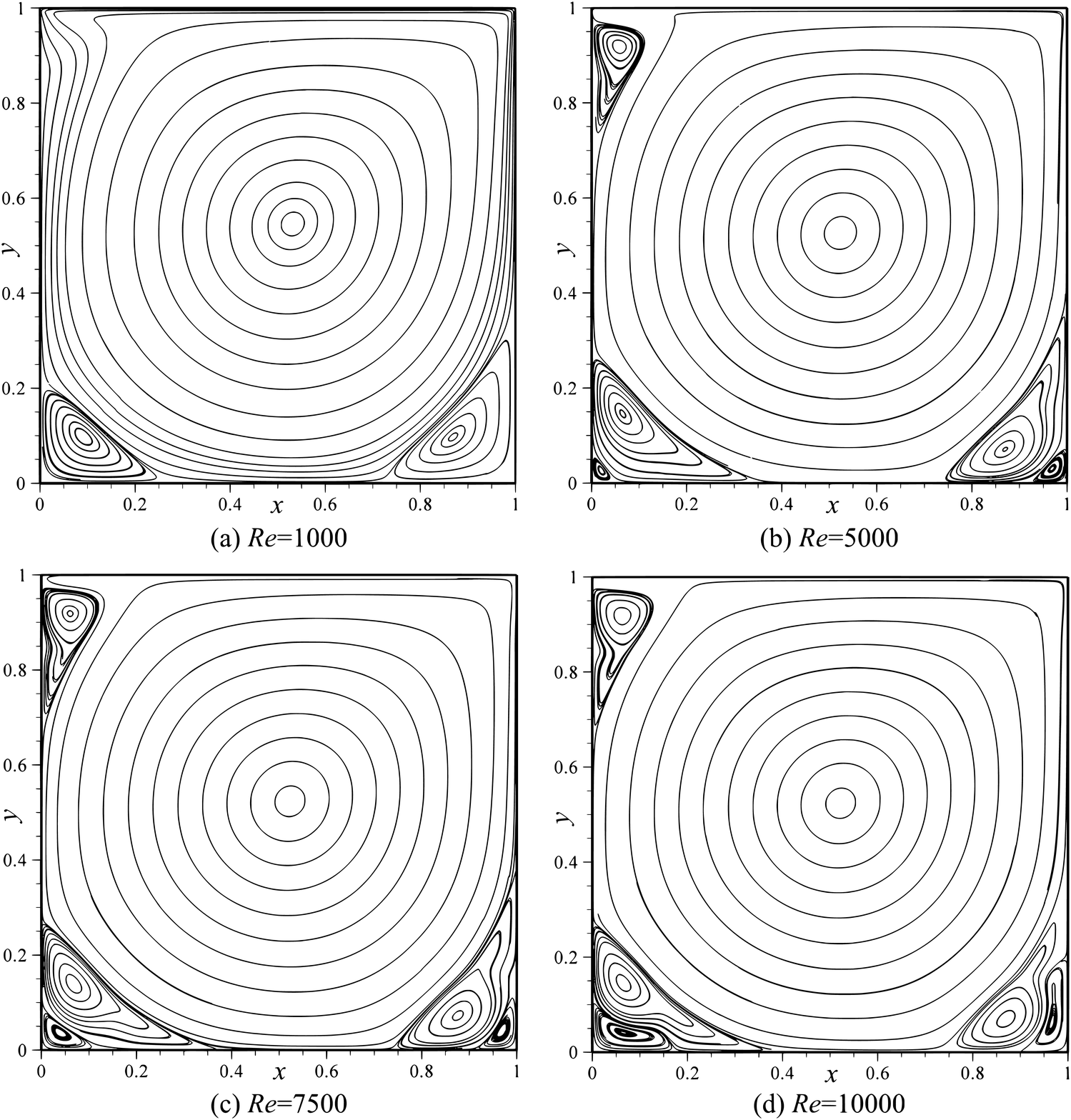}
  \caption{Streamline patterns for different Reynolds number calculated using the EMMS-based turbulence model.}\label{fig9}
\end{figure}

Fig. 9 shows the streamline patterns of our computed results in the cavity for the four different Reynolds numbers. As the Reynolds number increases, the primary, secondary and even tertiary corner vortices can be captured. When $Re$=1000, secondary vortices only appear in the lower left and right corners. When $Re$=5000, besides secondary vortices appearing at the same place, an additional secondary vortex also emerges in the upper left corner. Meanwhile, tertiary vortices can be observed in the lower left and right corners below the corresponding secondary vortices. When $Re$=7500 and 10000, similar phenomena to those for $Re$=5000 are still observed, except that the tertiary vortices become larger. However, when $Re$=7500 and 10000, the tertiary vortices become distorted compared with the reference solution (Ghia et al., 1982).

The standard $k$-$\varepsilon$ model was also used to determine this lid-driven cavity flow to provide another reference solution. The streamline patterns calculated by the two models at $Re=10000$ are compared in Fig. 10. The standard $k$-$\varepsilon$ model fails to predict the tertiary vortices in the lower left and right corners, whereas our proposed model successfully captures them. Specially, it should be noted that in current EMMS-based turbulence model zero-equation model was used to compute $\nu_{\mathrm{t}}$. Usually, zero-equation model is thought to be inferior to the standard $k$-$\varepsilon$ model, but here the results of the EMMS-based turbulence model are better than those of the standard $k$-$\varepsilon$ model, further illustrating the advantages of our work. The main reason for this is that the EMMS-based turbulence model includes turbulent structural parameters. Generally speaking, the standard $k$-$\varepsilon$ model regards the whole fluid in the cavity as turbulent state. However, near the walls especially the corners, the viscosity effect dominates the system rather than the inertia effect, so the fluid should be close to laminar flow instead of fully turbulent flow. It would obtain an unphysical solution in these sub-regions if the standard $k$-$\varepsilon$ model is used. In contrast, the EMMS-based turbulence model can treat it well due to that the flow everywhere has been considered as the coexistence of laminar and turbulent fluids, so the tertiary vortices in the lower left and right corners as well as the secondary vortex in the upper left corner can be captured successfully.

\subsubsection{Turbulent flow with forced convection in an empty room}

In this example, the standard $k$-$\varepsilon$ model was also revised, rather than a zero-equation model as in section 3.2.1.

The geometry used for this numerical example is shown in Fig. 11 with the following parameters: the width of the empty room was $H$=\,3.0\,m, the length of the empty room was $W$=\,$3H$, the inlet length of fluid was ${h_{\rm{in}}}$=\,0.056$ H$, which was located in the upper left corner of the empty room and the output length of fluid was ${h_{\rm{out}}}$=\,$0.16H$, which was located in the lower right corner of the empty room. Based on an inlet velocity of flow $U$=\,0.455\,m/s and the kinetic viscosity of fluid $\nu$=\,$1.53 \times {10^{ - 5}}\rm{{m^2}/s}$, the corresponding Reynolds number at the inlet was 5000. Nielsen's experimental data (Nielsen et al., 1978) was used as the reference solution.

\begin{figure}
  \centering
  \includegraphics[width=0.8\textwidth]{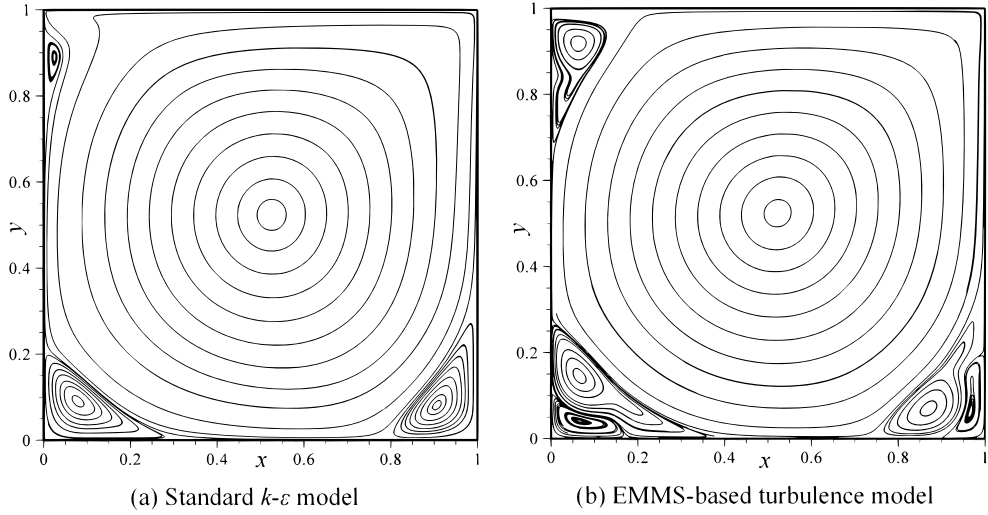}
  \caption{ Comparison of the streamline patterns predicted by the EMMS-based turbulence model and the standard $k$-$\varepsilon$ model at $Re$=10000.}\label{fig10}
\end{figure}

Fig. 12 shows the predicted flow patterns in the room with the standard $k$-$\varepsilon$ model and the EMMS-based turbulence model respectively. From Fig.12(a) we see that the secondary flow in the upper right and lower left corners can not be captured. However, from Fig.12(b) we can see that these two secondary flows are successfully predicted. The reason is that: near the walls especially the corners, the viscosity effect dominates the system rather than the inertia effect, so in these sub-regions we should not use the standard $k$-$\varepsilon$ model because it regards the whole fluid in the room as turbulent state; however, the simulation results suggest that the EMMS-based turbulence model has the ability to deal with this correctly.

\begin{figure}[htbp]
  \centering
  \includegraphics[width=0.7\textwidth]{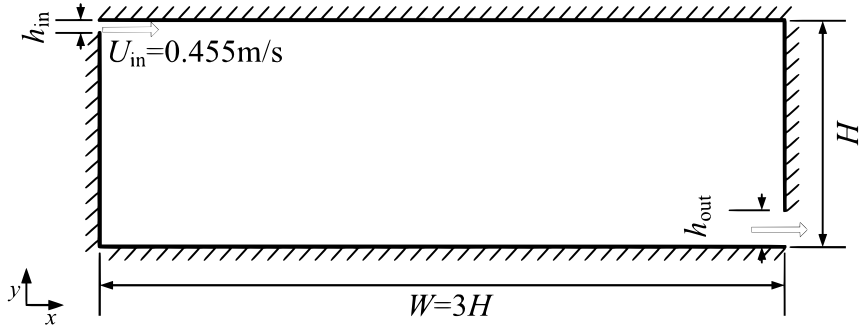}
  \caption{Geometry used for the forced convection example.}\label{fig11}
\end{figure}

The computed velocity profiles at vertical section of $x=H$ and horizontal section of $y=0.972H$ are presented in Fig. 13. Each profile is also compared with Nielsen's experimental data as well as the calculated results by using the standard $k$-$\varepsilon$ model. From Fig.13(a) we can see that the results of EMMS-based turbulence model are in good agreement with experimental data. Additionally, compared with the calculated results of the standard $k$-$\varepsilon$ model, there is a little accuracy improvement in the results of EMMS-based turbulence model. Further, from Fig.13(b) we can see that the results of EMMS-based turbulence model are closer to experimental data than those of the standard $k$-$\varepsilon$ model, and the accuracy improvement is obvious. Meanwhile, considering that the standard $k$-$\varepsilon$ model can not capture the secondary flow in the upper right and lower left corners while the EMMS-based turbulence model can do these (see in Fig.12), it can be concluded that our work is valid to improve the accuracy of turbulence modeling.

\begin{figure}[htbp]
  \centering
  \includegraphics[width=0.7\textwidth]{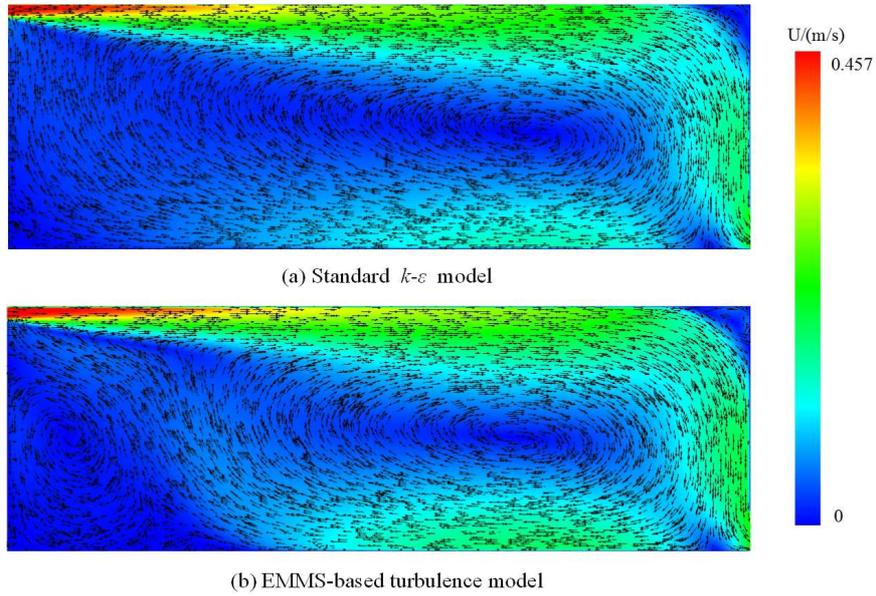}
  \caption{Predicted flow patterns in the room.}\label{fig12}
\end{figure}

\begin{figure}[htbp]
  \centering
  \includegraphics[width=0.85\textwidth]{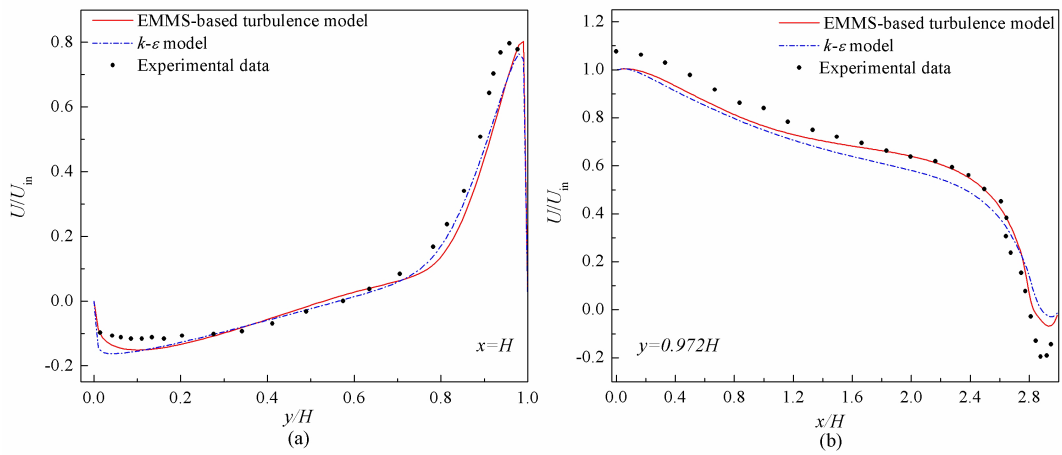}
  \caption{Comparison of computed velocity profile with experimental data at different sections: (a) $x=H$; (b) $y=0.972H$}\label{fig13}
\end{figure}

\section{Conclusions}
\label{}
We proposed an EMMS-based turbulence model in which single-phase flow is regarded as a mixture of turbulent and non-turbulent fluids, and the turbulence stability condition is quantified to close turbulent dynamic equations, allowing us to optimize the inhomogeneous structural parameters of turbulence. This meant that the corresponding turbulent viscosity coefficient could be constructed, improving the numerical simulation of turbulence. This is because that: the traditional turbulence models regard the whole fluid as fully turbulent state, but it is indeed unphysical near the walls especially the corners where the viscosity effect dominates, so the fluid flow in these sub-regions should be close to laminar flow rather than turbulent flow; in contrast, our EMMS-based turbulence model can deal with this successfully due to its physical basis revealing that complex flows can be viewed as a mixture of turbulent and non-turbulent fluids. To validate the effectiveness of the developed model, we used it to simulate two benchmark problems, lid-driven cavity flow and turbulent flow with forced convection in an empty room. The numerical results show that the developed model can indeed improve the accuracy of numerical simulation of turbulence and capture the detailed structure of turbulence such as secondary and tertiary vortices. This improvement is related to the model considering the governing principles of meso-scale structure.

However, the model still possesses some limitations. For example, the density of turbulent eddies was assumed to be $\rho_{\mathrm{\,e}}$=\,$0.99\rho_{\mathrm{\,l}}$, the superficial velocity of the laminar component $U_{\rm{l}}$ was fixed at 0.001 m/s, the rotation and interaction of turbulent eddies were not considered, and the heterogeneous structural parameter $d_{\mathrm{e}}$ was not included. We will attempt to address these problems to further improve the EMMS-based turbulence model in the future.

\begin{thenomenclature}
\small
 \nomgroup{A}
  \item [{\makebox[1cm][l]{$\bar e(\lambda)$}}]\begingroup mean kinetic energy, J/(kg{\kern 1pt}s)\nomeqref {40}
		\nompageref{34}
  \item [{\makebox[1cm][l]{$\mathbf{F}$}}]\begingroup resistant force, kg{\kern 1pt}m/$\rm {s}^2$\nomeqref {40}
		\nompageref{34}
  \item [{\makebox[1cm][l]{$\mathbf{g}$}}]\begingroup gravitational acceleration, m/$\rm {s}^2$\nomeqref {40}
		\nompageref{34}
  \item [{\makebox[1cm][l]{$C_{\rm{D}}$}}]\begingroup drag coefficient\nomeqref {40}
		\nompageref{34}
  \item [{\makebox[1cm][l]{$d_{\rm{e}}$}}]\begingroup equivalent diameter of turbulent eddies, m\nomeqref {40}
		\nompageref{34}
  \item [{\makebox[1cm][l]{$E_{\rm{DIS}}$}}]\begingroup viscous dissipation inside the drop/bubble during its breakage, J/(kg{\kern 1pt}s)\nomeqref {40}
		\nompageref{34}
  \item [{\makebox[1cm][l]{$f$}}]\begingroup volume fraction of turbulent eddies\nomeqref {40}
		\nompageref{34}
  \item [{\makebox[1cm][l]{$f_{\rm{BV}}$}}]\begingroup volume ratio of a smaller turbulent eddy produced by breakage of a large one\nomeqref {40}
		\nompageref{34}
  \item [{\makebox[1cm][l]{$k$}}]\begingroup wave number\nomeqref {40}\nompageref{34}
  \item [{\makebox[1cm][l]{$L$}}]\begingroup mixing length, m\nomeqref {40}
		\nompageref{34}
  \item [{\makebox[1cm][l]{$n_{\rm{e}}$}}]\begingroup number density of initial turbulent eddies of size $d_{\rm{e}}$\nomeqref {40}
		\nompageref{34}
  \item [{\makebox[1cm][l]{$P_{\rm{1}}$}}]\begingroup pressure at the top of reactor, kg/(m{\kern 1pt}$\rm {s}^2$)\nomeqref {40}
		\nompageref{34}
  \item [{\makebox[1cm][l]{$P_{\rm{2}}$}}]\begingroup pressure at the bottom of reactor, kg/(m{\kern 1pt}$\rm {s}^2$)\nomeqref {40}
		\nompageref{34}
  \item [{\makebox[1cm][l]{$P_{\rm{e}}$}}]\begingroup probability of a current turbulent eddy breaking into two smaller turbulent eddies\nomeqref {40}
		\nompageref{34}
  \item [{\makebox[1cm][l]{$r$}}]\begingroup radius of curvature of turbulent eddies, m\nomeqref {40}
		\nompageref{34}
  \item [{\makebox[1cm][l]{$Re$}}]\begingroup Reynolds number\nomeqref {40}
		\nompageref{34}
  \item [{\makebox[1cm][l]{$U$}}]\begingroup local mean velocity, m/s\nomeqref {40}
		\nompageref{34}
  \item [{\makebox[1cm][l]{$U_{\rm{e}}$}}]\begingroup superficial velocity of turbulent eddies, m/s\nomeqref {40}
		\nompageref{34}
  \item [{\makebox[1cm][l]{$U_{\rm{g}}$}}]\begingroup superficial velocity of bubbles, m/s\nomeqref {40}
		\nompageref{34}
  \item [{\makebox[1cm][l]{$U_{\rm{in}}$}}]\begingroup inlet velocity, m/s\nomeqref {40}
		\nompageref{34}
  \item [{\makebox[1cm][l]{$U_{\rm{l}}$}}]\begingroup superficial velocity of laminar flow, m/s\nomeqref {40}
		\nompageref{34}
  \item [{\makebox[1cm][l]{$u_{\rm}$}}]\begingroup slip velocity between the turbulent eddies and laminar fluid, m/s\nomeqref {40}
		\nompageref{34}
  \item [{\makebox[1cm][l]{$W_{\rm{\nu}}$}}]\begingroup energy consumption of molecular viscosity per unit mass and per unit time, J/(kg{\kern 1pt}s)\nomeqref {40}
		\nompageref{34}
  \item [{\makebox[1cm][l]{$W_{\rm{st}}$}}]\begingroup energy stored in the energy-containing eddies per unit mass and per unit time, J/(kg{\kern 1pt}s)\nomeqref {40}
		\nompageref{34}
  \item [{\makebox[1cm][l]{$W_{\rm{te}}$}}]\begingroup turbulent dissipation per unit mass and per unit time, J/(kg{\kern 1pt}s)\nomeqref {40}
		\nompageref{34}
  \item [{\makebox[1cm][l]{$W_{\rm{T}}$}}]\begingroup total energy consumption per unit mass and per unit time, J/(kg{\kern 1pt}s)\nomeqref {40}
		\nompageref{34}
  \item [{\makebox[1cm][l]{${\bar u_{\rm{\lambda}}}$}}]\begingroup turbulent velocity of turbulent eddies of size $\lambda$, m/s\nomeqref {40}
		\nompageref{34}
  \item [{\makebox[1cm][l]{${\dot n_{\rm{\lambda }}}$}}]\begingroup number density of turbulent eddies of size $\lambda$\nomeqref {40}
		\nompageref{34}

 \nomgroup{G}

  \item [{\makebox[0.8cm][l]{$\eta$}}]\begingroup Kolmogorov scale, m\nomeqref {40}
		\nompageref{34}
  \item [{\makebox[0.8cm][l]{$\eta_{\rm{_D}}$}}]\begingroup drop/bubble viscosity, $\rm{m}^2$/s\nomeqref {40}
		\nompageref{34}
  \item [{\makebox[0.8cm][l]{$\lambda_{\rm{min}}$}}]\begingroup minimum size of the inertial subrange, m\nomeqref {40}
		\nompageref{34}
  \item [{\makebox[0.8cm][l]{$\mu_{\rm}$}}]\begingroup viscosity coefficient of the laminar component of turbulence, kg/(m{\kern 1pt}s)\nomeqref {40}
		\nompageref{34}
  \item [{\makebox[0.8cm][l]{$\nu$}}]\begingroup kinematic viscosity, $\rm{m}^2$/s\nomeqref {40}
		\nompageref{34}
  \item [{\makebox[0.8cm][l]{$\nu_{\rm{0}}$}}]\begingroup intrinsic kinematic viscosity of the fluid, $\rm{m}^2$/s\nomeqref {40}
		\nompageref{34}
  \item [{\makebox[0.8cm][l]{$\nu_{\rm{eff}}$}}]\begingroup efficient kinematic viscosity, $\rm{m}^2$/s\nomeqref {40}
		\nompageref{34}
  \item [{\makebox[0.8cm][l]{$\nu_{\rm{t}}$}}]\begingroup turbulent kinematic viscosity, $\rm{m}^2$/s\nomeqref {40}
		\nompageref{34}
  \item [{\makebox[0.8cm][l]{$\omega_{\rm{e,\kern 1pt \lambda}}$}}]\begingroup arrival frequency of turbulent eddies\nomeqref {40}
		\nompageref{34}
  \item [{\makebox[0.8cm][l]{$\rho_{\rm{e}}$}}]\begingroup density of turbulent eddies, kg/$\rm{m}^3$\nomeqref {40}
		\nompageref{34}
  \item [{\makebox[0.8cm][l]{$\rho_{\rm{l}}$}}]\begingroup density of laminar flow, kg/$\rm{m}^3$\nomeqref {40}
		\nompageref{34}
  \item [{\makebox[0.8cm][l]{$\sigma$}}]\begingroup interfacial tension of turbulent eddies, $\rm {s}^{-1}$\nomeqref {40}
		\nompageref{34}
  \item [{\makebox[0.8cm][l]{$\tau_{\rm{_D}}$}}]\begingroup viscous stress inside a breaking drop/bubble, $\rm {s}^{-1}$\nomeqref {40}
		\nompageref{34}

 \nomgroup{S}

  \item [{\makebox[0.8cm][l]{$\rm{\lambda}$}}]\begingroup size of turbulence eddies\nomeqref {40}
		\nompageref{34}
  \item [{\makebox[0.8cm][l]{$\rm{eff}$}}]\begingroup effective\nomeqref {40}
		\nompageref{34}
  \item [{\makebox[0.8cm][l]{$\rm{e}$}}]\begingroup eddies\nomeqref {40}\nompageref{34}
  \item [{\makebox[0.8cm][l]{$\rm{l}$}}]\begingroup laminar\nomeqref {40}\nompageref{34}
  \item [{\makebox[0.8cm][l]{$\rm{t}$}}]\begingroup turbulent\nomeqref {40}\nompageref{34}

 \nomgroup{Z}

  \item [{\makebox[2.5cm][l]{CFD}}]\begingroup Computational Fluid Dynamics\nomeqref {40}
		\nompageref{34}
  \item [{\makebox[2.5cm][l]{DNS}}]\begingroup Direct Numerical Simulation\nomeqref {40}
		\nompageref{34}
  \item [{\makebox[2.5cm][l]{EMMS}}]\begingroup Energy-Minimization Multi-Scale\nomeqref {40}
		\nompageref{34}
  \item [{\makebox[2.5cm][l]{LES}}]\begingroup Large Eddy Simulation\nomeqref {40}
		\nompageref{34}
  \item [{\makebox[2.5cm][l]{OpenFOAM}}]\begingroup Open Source Field Operation And Manipulation\nomeqref {40}
		\nompageref{34}
  \item [{\makebox[2.5cm][l]{PISO}}]\begingroup Pressure Implicit with Splitting of Operators\nomeqref {40}
		\nompageref{34}
  \item [{\makebox[2.5cm][l]{RANS}}]\begingroup Reynolds-Averaged Navier-Stokes\nomeqref {40}
		\nompageref{34}

\end{thenomenclature}

\section*{Acknowledgement}
\label{}
This work was financially supported by the National Natural Science Foundation of China (No. 21106155), Science Foundation of the Chinese Academy of Sciences (No. XDA07080303) and China Postdoctoral Science Foundation (No. 2012M520385).

\bibliographystyle{elsarticle-harv}



\end{document}